\documentclass[useAMS,usenatbib]{mn2e}

\usepackage{graphicx}
\usepackage{amsmath}
\usepackage{amssymb}

\newcommand{\bm}[1]{\boldsymbol{#1}}

\title[Thermal Conduction in the Virgo Cluster]{The Effects of Thermal
Conduction on the ICM of the Virgo Cluster}

\author[E.C.D. Pope, G. Pavlovski, C.R. Kaiser, H. Fangohr] {Edward
C.D. Pope$^{1,2}$\thanks{E-mail:edpope@soton.ac.uk}, Georgi
Pavlovski$^{1}$, Christian R. Kaiser$^{1}$, Hans Fangohr$^{2}$ \\
$^{1}$School of Physics \& Astronomy, University of Southampton, UK,
SO17 1BJ\\ $^{2}$School of Engineering Sciences, University of
Southampton, UK, SO17 1BJ}

\begin{document}

\pagerange{\pageref{firstpage}--\pageref{lastpage}} \pubyear{2005}

 \maketitle

\label{firstpage}

\begin{abstract}

Thermal conduction has been suggested as a possible mechanism by which
sufficient energy is supplied to the central regions of galaxy
clusters to balance the effect of radiative cooling. Recent high
resolution observations of the nearby Virgo cluster make it an ideal
subject for an attempt to reproduce the properties of the cluster by
numerical simulations, since most of the defining parameters are
comparatively well known. Here we present the results of a simulated
high-resolution, 3-d Virgo cluster for different values of thermal
conductivity (1, 1/10, 1/100, 0 times the full Spitzer
value). Starting from an initially isothermal cluster atmosphere we
allow the cluster to evolve freely over timescales of roughly $
1.3-4.7 \times 10^{9} $ yrs. Our results show that thermal
conductivity at the Spitzer value can increase the central ICM
radiative cooling time by a factor of roughly 3.6. In addition, for
larger values of thermal conductvity the simulated temperature and
density profiles match the observations significantly better than for
the lower values. However, no physically meaningful value of thermal
conductivity was able to postpone the cooling catastrophe
(characterised by a rapid increase in the central density) for longer
than a fraction of the Hubble time nor explain the absence of a strong
cooling flow in the Virgo cluster today. We also calculate the
effective adiabatic index of the cluster gas for both simulation and
observational data and compare the values with theoretical
expectations. Using this method it appears that the Virgo cluster is
being heated in the cluster centre by a mechanism other than thermal
conductivity. Based on this and our simulations it is also likely that
the thermal conductvity is suppressed by a factor of at least 10 and
probably more. Thus, we suggest that thermal conductvity, if present
at all, has the effect of slowing down the evolution of the ICM, by
radiative cooling, but only by a factor of a few.

\end{abstract}


\section{Introduction} \label{sIntro}

The strong central peak of the X-ray surface brightness profiles of
many clusters of galaxies is generally interpreted as the signature of
a cooling flow \citep{cowie77, fab94}. Standard cooling flow theory
suggests that this peak should coincide with the deposition of large
amounts of cold gas and significant excess of star formation at the
centre of the cooling flow. However, the XMM-Newton and Chandra
satellites have demonstrated both the lack of the expected cold gas,
at temperatures below roughly 1 keV \citep[e.g.][]{edge01}, and that
spectroscopically measured mass-deposition rates are roughly one tenth
of the value inferred from the peaks of cluster X-ray surface
brightness maps \citep{voigt04}. These findings suggest that the gas
in the cluster cores is reheated; either continually or periodically,
in order to produce the observed minimum central temperature (or
entropy) and the lack of significant excess star formation. In recent
years, work has concentrated on two main possibilities for heating the
gas in the central regions of clusters: (1) heating by outflows from
AGN \citep[e.g.][]{tabor93,bub01,nature} and (2) transportation of
heat from the outer regions of the cluster by thermal conduction
\citep[e.g.][]{zakamska03, voigt02, voigt04}. In this paper we shall
concentrate purely on thermal conduction.

The presence of large temperature gradients in the centres of many
relaxed clusters allows the possibility that thermal conduction may
play a role in reducing the radiative energy losses in the cluster
centre. Indeed, \cite{zakamska03} show that for several galaxy
clusters the inward heatflow from the outer regions due to thermal
conduction may be sufficient to balance the radiative losses although,
in a different sample \cite{voigt02} show that thermal conduction is
only able to balance radiative losses in the outer regions of the
cluster core. However, though it is possible to achieve an energy
balance for these clusters in their present state, this requires the
thermal conductivity to vary as a function of radius. Furthermore, it
is unclear what effect thermal conduction would have on the intra
cluster medium (ICM) if it constitutes a significant process
throughout the lifetime of the cluster. Since thermal conduction acts
to oppose the formation of the temperature gradient that causes it,
one might expect that the temperature and density profiles for those
clusters in which thermal conduction plays a significant role to
appear different compared to those clusters in which thermal
conduction is less effective. In particular, the temperature profiles
in such clusters could be much flatter than in clusters where thermal
condcution is heavily suppressed. In addition, temperature gradients
also occur outwards towards the edge of clusters. Therefore, it is
possible that the high thermal conductivities required to balance the
radiative losses may also have a cooling effect by transporting energy
away from the cluster centre and out into intercluster space
\citep{loeb02}.

One of the primary problems for studying the effects of thermal
conduction in plasmas is the unknown value of the thermal
conductivity. The theoretical value for the thermal conductivity of a
fully ionized, unmagnetized plasma was originally calculated by
\cite{spitzer}. However, observations indicate the existence of a
magnetic field in the ICM \citep[B $ < $ 10 $\mu $G;
e.g.][]{carilli02} which can affect this value dramatically. For
magnetic fields of these strengths the electron and proton gyro radii
are much smaller than their mean free paths along the field lines. The
precise effect that magnetic fields have on thermal conductivity is
unclear, although for magnetic fields of the observed strengths the
effective thermal conductivity is determined by the topology of the
field \citep{markevitch03}. A magnetic field tangled on scales much
longer than the electron mean free path will suppress thermal
conduction perpendicular to the field lines reducing the thermal
conductivity to at most one third of the full Spitzer value (if
transport along the field line is unaltered) \citep{sarazin}. For a
magnetic field tangled on scales shorter than the electron mean free
path the conductivity is suppressed by a factor of up to 100 as the
electrons have to travel further to diffuse a given distance
perpendicular to the field \citep{tribble89}. In addition, for
tangling lengths that are comparable to the electron mean free path,
conduction along the magnetic field line can be reduced by a factor of
roughly 10 by the effect of magnetic mirrors
\citep{malyshkin01}. Alternatively, recent theoretical work by
\cite{narmed01} and \cite{cho03} has shown that for certain spectra of
field fluctuations, turbulent magnetic fields are less efficient at
suppressing thermal conduction than previously thought. The highly
tangled, turbulent magnetic field may allow significant cross field
diffusion such that the thermal conductivity remains of the order of
the Spitzer value. In any case the collisional thermal conductivity on
100 kpc scales is likely to be reduced by a factor of 3-100
\citep{markevitch03}.

We define the suppression factor, $f$, by $ \kappa = f \kappa_{\rm S}
$, where $ \kappa $ is the actual thermal conductivity and $
\kappa_{\rm S} $ is the Spitzer thermal conductivity. To date this
factor has been very difficult to constrain from observations:
\cite{voigt02} and \cite{zakamska03} find that suppression factors of
$ > $ 0.3 are sufficient to balance radiative losses in some cluster
while other cluster require unphysically high values ($f$ $ > $ 1). In
other cases, the temperature drops across cold fronts in A2142
\citep{ettori00} implies a suppression factor of $ < $ 0.004. At the
boundaries between the ICM and galaxy-size dense gas clouds in Coma
\citep{Vik2001} the thermal conductivity is found to be suppressed by
a factor of order one hundred. However, one might reasonably expect
both of these cases not to be characteristic of the bulk ICM, since
they contain boundaries between different gas phases and disjointed
magnetic fields. Another estimate uses the existence of cold filaments
in galaxy clusters \citep{NB04} to put constraints on the value of the
suppression factor. Although not all of the values of the required
quantities are well known, they find that for the Perseus cluster the
suppresion factor is $ < $ 0.04, which result they consider to be
compatible with the results of \cite{markevitch03} based on
observations of temperature gradients present in A754.

The aim of the work presented here is to investigate whether thermal
conduction can prevent the catastrophic radiative cooling of the gas
at the centres of galaxy clusters by transporting thermal energy from
the cluster outskirts to the centre. We also investigate whether this
process gives rise to gas temperature and density profiles compatible
with those derived from X-ray observations.

The plan of this paper is as follows. In Section \ref{prop} we discuss
why thermal conduction would be an unlikely solution to the cooling
flow problem from analytical predictions. Section \ref{sect:model},
describes the numerical model and code which we use to obtain our
results. Section \ref{init} states the initial conditions specific to
the problem we are investigating. Section \ref{sims} gives the details
of all simulations, e.g. time simulated and value of thermal
conductivity used. In Section 4 we present the results of our
simulations in the form of temperature and density profiles,
emissivity profiles, the effective adiabatic index, central
temperature and density, and mass flow rates respectively. We also
discuss the comparisons of these results with recent observations of
the Virgo cluster and their implications. We present our conclusions
in Section \ref{conc} and demonstrate the numerical convergence of our
results in the Appendix.

\section{Heating by Thermal Conduction} \label{prop}

Thermal conduction transfers heat so as to oppose the temperature
gradient which causes the transfer. Therefore, if thermal conduction
occurs at all in galaxy clusters, it will certainly result in some
heating of the cluster centre, but to what extent? If thermal
conduction is to provide a solution to the cooling flow problem it
must fulfill the following conditions:
\begin{enumerate}
\item temperature and density profiles must be simultaneously
comparable with observations and stable for cosmologically significant
timescales \citep[e.g.][]{allen01}.
\item the deposition of large amount of cold gas in the centre of the
cluster must be suppressed \citep[e.g.][]{edge01}.
\end{enumerate}

In order to assess the likelyhood that thermal conduction is capable
of satisfying the above requirements we make analytical predictions
based on the relevant hydrodynamic equations.

The rate of radiative energy loss per unit volume ($ \epsilon_{\rm
rad} $) is given by equation (\ref{eq:radloss}) where $n_{\rm e}$ is
the electron concentration,$ n_{p}$ is the concentration of
protons/positive ions and $ \Lambda_{\rm rad}(T,Z) $ is the cooling
function

\begin{equation} \label{eq:radloss}
\epsilon_{\rm rad} = -n_{\rm e}n_{\rm p} \Lambda_{\rm rad}(T,Z) ~[{\rm erg
~s^{-1} cm^{-3}}].
\end{equation}

The rate of heating due to thermal conduction is given by the
divergence of the thermal flux

\begin{equation} \label{eq:etherm}
\epsilon_{\rm cond} = \frac{1}{r^{2}} \frac{\rm d}{{\rm d}r}\bigg
(r^{2} \kappa \frac{{\rm d}T}{{\rm d} r} \bigg) ~[{\rm erg ~s^{-1}
cm^{-3}}],
\end{equation}
where $ \kappa $ is the thermal conductivity defined in the
introduction.  

For the high temperatures of cluster atmospheres the coefficients of
thermal conductivity and viscosity are given by the value calulated by
\cite{spitzer}. In the absence of magnetic fields the coefficients of
Spitzer thermal conductivity and viscosity are given by

\begin{equation} \label{eq:therm}
\kappa_{\rm s} = \frac{1.84 \times 10^{-5}}{\ln{\Lambda_{\rm
c}}}T^{5/2} ~[{\rm erg ~s^{-1}cm^{-1}K^{-1}}]
\end{equation}

\begin{equation} \label{eq:visc}
\mu_{\rm s} = \frac{2.5 \times 10^{-15}}{\ln{\Lambda_{\rm c}}}
T^{5/2} ~[{\rm g~s^{-1}cm^{-1}}],
\end{equation}
where $ \ln {\Lambda_{\rm c}} $ is the Coulomb logarithm and $
\Lambda_{\rm c} $ is given by equation (\ref{eq:coul}).

Due to the large variation in density over cluster-scale distances it
is necessary to take into account the variation of the Coulomb
logarithm given by \citep[e.g.][]{plasmas}

\begin{equation} \label{eq:coul}
\Lambda_{\rm c} = 24 \pi n_{\rm e} \bigg ({\frac{8 \pi e^{2} n_{\rm
e}}{k_{\rm B}T}} \bigg) ^{-3/2}.
\end{equation}

The opposing energy fluxes of radiative cooling and thermal conduction
determine the behaviour of the cluster gas. For a spherically
symmetric cooling flow, in which we ignore the effect of viscous
heating and gravitational effects, the energy equation is
\begin{equation}\label{eq:energy}
\frac{\rm d}{{\rm d}t}\Bigg(\frac{3k_{\rm B} n_{\rm e}T}{2\mu m_{\rm
    p}}\Bigg) = -\epsilon_{\rm rad}+\epsilon_{\rm cond} ~[{\rm erg
    ~s^{-1}cm^{-3}}],
\end{equation}
where the symbols have the definitions given above.

A rough estimate of the cooling lifetimes of the ICM can be used to
illustrate the point that thermal conduction would be an unlikely
solution to the cooling flow problem. For the case in which thermal
conduction is completely supressed the cooling time of the gas can be
defined as

\begin{equation} \label{eq:tcool}
 t_{\rm cool} \approx \frac{n_{\rm e}k_{\rm B}T}{n_{\rm e}^2
 \Lambda_{\rm rad}}~[{\rm s}],
\end{equation}
where $ \Lambda_{\rm rad} = 2.1 \times 10^{-27}\sqrt {T} $ erg
s$^{-1}{}$cm$^{-3} $ \citep{rybicki} is the cooling function for pure
bremsstrahlung emission, $T$ is the gas temperature and $n_{\rm e}$ is
the electron number density.

Using equation (\ref{eq:tcool}), with the initial conditions near the
centre of the Virgo cluster (see section \ref{init}), the central
cooling time is approximately 200 Myr. This is comparable to the
results of our simulation without thermal conduction. The cooling time
at $\sim $ 100 kpc is $ \sim 10^{12} $ yrs.

Given the form of equation (\ref{eq:tcool}), the central cooling time
for a radiatively cooling and thermally conducting ICM is given by

\begin{equation} \label{eq:tcooltot}
 t_{\rm cool} \approx \frac{n_{\rm e}k_{\rm B}T}{\epsilon_{\rm rad}-
 \epsilon_{\rm cond}}~[{\rm s}],
\end{equation}
where $\epsilon_{\rm rad}$ and $ \epsilon_{\rm cond} $ are defined by
equations (\ref{eq:radloss}) and (\ref{eq:etherm}) respectively.

Thus, if thermal conductivity is the sole process by which the
radiative cooling is opposed, then the predicted cooling time of 200
Myr, based on observations, must be extended by at least an order of
magnitude. From equation (\ref{eq:tcooltot}) this requires the energy
flux due to thermal conduction to be equal to the radiative energy
flux to within better than 1 \%. Such a situation may be possible for
exceptional clusters e.g. those with very low density, so that the
radiative losses are small, and high temperatures, so that the thermal
conductivity is large. However, due to the independent nature of the
competing processes it seems unfeasible to expect this to be the case
in every galaxy cluster. In addition, although thermal conductivity
probably increases the cooling time of the gas at the cluster centre
by a factor of a few, in doing so it also reduces the cooling time at
larger radii introducing a secondary cooling problem.

However, we note that the above analysis is based only on estimates,
since it is impossible to solve the full hydrodynamic equations
analytically. Therefore, in order to better understand the effect of
thermal conduction on the ICM we perform 3-d simulations in which the
hydrodynamic equations are solved numerically.

\section{Numerical Model} 

\subsection{The Code}
\label{sect:model}

We performed high resolution, 3-d numerical hydrodynamic simulations
to produce temperature and density profiles of model Virgo
clusters. For simplicity we assume that the dark matter halo of our
simulated cluster is fully formed and stationary throughout the
simulation. We also neglect the self-gravity of the gas. Thus the
gravitational potential confining the gas is fixed. We also
self-consistently include the effects of radiative cooling, thermal
conductivity and viscosity. We do not include any external heating
mechanisms such as AGN. By neglecting the self-gravity of the cluster
gas we are able to run the simulations over cosmologically relevant
timescales (${\rm few} \times 10^{8} - {\rm few} \times 10^{9} $ yrs).

Our simulations were performed with FLASH2.3, an Adaptive Mesh
Refinement (AMR) hydrodynamical code developed and made public by the
ASCII Center at the University of Chicago \citep{fryxell}. FLASH is a
modular block-structured AMR code, parallelised using the Message
Passing Interface (MPI) library.

FLASH solves the Riemann problem on a cartesian grid using the
Piecewise-Parabolic Method \citep[PPM;][]{woodward}. It uses a
criterion based on the dimensionless second derivative $ D^{2} \equiv
F{\rm d}^{2}F/{\rm d}x^{2} / ({\rm d}F/{\rm d}x)^{2} $ of a fluid
variable $F$ to increase the resolution adaptively whenever $ D^{2} >
c_{1} $ and de-refine the grid when $ D^{2} < c_{2} $, $ c_{1,2} $ are
tolerance parameters. When a region requires refining, child cells of
half the size of the parent cells are placed over the offending
region, and the coarse solution is interpolated. In our simulations we
have used density, pressure and temperature as the refinement fluid
variable $F$ \citep[see also][]{vecchia04}.

FLASH imposes the analytic gravitational potential to the grid cell,
and computes the corresponding gravitational acceleration. This is
done only once since we neglect the self-gravity of the gas. To
interpolate the intial density field to the FLASH mesh, we impose a
uniform initial grid with a sphere of higher resolution in the central
regions with a radius of 16.2 kpc, because of the steeper gradients in
the cluster centre. This is the minimum refinement allowed during the
simulation, the maximum refinement was set to 9 refinement levels,
equivalent to $2048^{3}$ uniform computational cells. We use periodic
boundary conditions and a computational box size large enough (648
kpc) that the density at the cluster edge remains approximately
constant over the lifetime of the simulation.

FLASH calculates three limiting timesteps based on the requirements
for consistency in the hydrodynamics, radiative cooling and diffusive
processes, i.e. thermal conduction and viscosity, of the solution in
the entire computational grid. The solution is then advanced by
choosing the minimum of these timesteps. The hydro timestep is derived
by the Courant condition $ {\rm d}t = C \delta x /c_{\rm s} $ where
$\delta x $ is the cell dimension, $ c_{\rm s} $ is the sound speed in
that cell and $C$ is the CFL coeffient which is set to 0.8 in our
simulations.  The cooling timestep is given by: $ {\rm d}t = ei/ ({\rm
d}ei/{\rm d}t) $ where $ ei $ is the internal energy per unit volume
and ${\rm d}ei/{\rm d}t $ is the rate of change of internal energy. We
usually find this timestep to be very large except for very high
cooling rates. The diffusion timestep is given by: $ {\rm d}t = 0.5
({\rm min} ({\rm d}x^{2}))/{\rm max}(\chi) $ where $ {\rm min}({\rm
d}x^{2}) $ is the smallest cell dimension, $\chi $ is the diffusivity
and $ {\rm max}(\chi)$ is the largest diffusivity in the computational
zone. This timestep is smaller for larger values of thermal conduction
and viscosity.

The simulations were performed using Iridis, one of the University of
Southampton`s Beowulf clusters. Iridis has 115 computational nodes
connected with myrinet of which we usually used 8 to 10 nodes (16 to
20 processors).

\subsection{Initial Conditions} \label{init}

From the assumption that the Virgo cluster is currently approximately
in hydrostatic equilibrium it is possible to derive the gravitational
acceleration as a function of radius within the cluster if both the
temperature and the density distributions of the gas are well
defined. For these we use the best-fit functions for the temperature
and electron number density distributions from \cite{ghizzardi04} who
combine the data from Chandra, XMM-Newton and Beppo-SAX. They fit the
deprojected temperature and electron number density profiles with a
Gaussian (see equation \ref{eq:Tvirg}) and double beta profile (see
equation \ref{eq:virgdens}) respectively. We consider this method more
accurate than estimating the appropriate parameters for a Navarro,
Frenk \& White potential \citep{nfw97}, since the gravitational
potential in the central regions of a real galaxy cluster will be
dominated by the central galaxy.

\begin{equation} \label{eq:Tvirg}
 T = T_{\rm 0} - T_{\rm
1}\exp\bigg(-\frac{1}{2}\frac{r^2}{\sigma^2}\bigg)[\rm ~K],
\end{equation}
the best fit values are $ T_{\rm 0} $= 2.78 $\times $10$^{7} $ K , $
T_{\rm 1} $ = 8.997 $\times $ 10$^{6}$ K and $ \sigma $ = 7.39$ \times
$10$^{22}$ cm.  

The electron density is given by:
\begin{equation} \label{eq:virgdens}
 n_{\rm e} = \frac{n_{\rm 0}}{(1+(r/r_{\rm 0})^2)^{\alpha_{{\rm 0}}}}
+ \frac{ n_{\rm 1}}{(1+(r/r_{\rm 1})^2)^{\alpha_{{\rm 1}}}}[\rm ~cm^{-3}],
\end{equation}
the best fit values are given by $ n_{\rm 0} $ = 0.089 cm$^{-3} $ , $
n_{\rm 1} $ = 0.019 cm$^{-3} $, $ r_{\rm 0} $ = 1.6 $ \times $10$^{22}
$cm, $ r_{\rm 1} $ = 7.39 $ \times $10$^{22}$ cm, $ \alpha_{{\rm 0}}$
= 1.52, $ \alpha_{{\rm 1}} $ = 0.705. 

The double $ \beta $-profile for the gas density found by
\cite{ghizzardi04} is interesting given the observation that although
in general cluster temperature profiles are consistent with a single
phase gas, the X-ray surface brightness is less centrally peaked than
expected. This is taken as evidence for distributed mass deposition
throughout the flow, caused by a multiphase gas \citep[][ and
references therein]{missing02}.

Work by \cite{evrard90} and \cite{nfw95} showed that the central
megaparsec of clusters are roughly isothermal due to the effect of
shock heating of the infall of the gas in the formation of the
cluster. Therefore our initial conditions assume that the ICM is
uniformly heated to the cluster Virial temperature and has an initial
density profile that ensures hydrostatic equilibrium with the
gravitational potential derived above. The assumption of a precisely
uniform initial temperature is adequate for the purposes of this work
since we simulate only the central few hundred kiloparsecs of the
Virgo cluster. However, we note that it may still be simplisitic with
regard to residual temperature and density perturbations arising from
the formation of the cluster which should be taken in to account for a
complete study of ICM evolution. However, this will be dealt with in
future work. 

We estimate the Virial temperature ($ T_{\rm vir} $) of the Virgo
cluster from

\begin{equation} \label{eq:Tvir}
T_{\rm vir} \approx \frac{G M_{\rm vir} \mu m_{\rm p}}{k_{\rm B} R_{\rm
vir}}~[K],
\end{equation}
where $ M_{\rm vir} $ is the Virial mass, $ R_{\rm vir} $ is the
Virial radius, $G$ is the Universal gravitational constant, and $ \mu
m_{\rm p} $ is the mean molecular mass of the gas.

We are able to estimate the Virial mass and radius using the
gravitational mass profile given by \cite{ghizzardi04}. We adopt a
Virial mass of $ 10^{13} \rm M_{\odot} $ with a Virial radius of 100
kpc and find the Virial temperature of the Virgo cluster to be $ \sim
3.1 \times 10^{7} $ K. The initial value for the uniform temperature
was determined by finding a temperature near to the Virial
temperature, for which the simulated temperature profile, after a
period of $ 10^{8} $ yrs of evolution of the ICM, was comparable to
current observations. We found that an initial temperature of $ 3
\times 10^{7} $ K resulted in an adequate temperature profile.

The initial density profile required for hydrostatic equilibrium is
determined by the derived gravitational potential and the initially
uniform temperature up to a multiplicative constant. This constant is
defined as the density at the cluster centre. Since the gas density in
the cluster outskirts will not vary much throughout the simulated
time, we adjust this constant such that the density in the outer
regions of our cluster is comparable to the currently observed density
in these localities. Thus we found an appropriate central density of $
5.15 \times 10^{-26} \rm g~ cm^{-3}$.

In our model, the ICM is a single, ideal fluid, of half-solar
metallicity which is assumed to be fully ionised with an adiabatic
index of 5/3. For such a composition the total mass density of the ICM
is given by: $ \rho = (8/7) m_{\rm p} n_{\rm e} $ where $ n_{\rm e} $
is the electron number density and $ m_{\rm p} $ is the proton mass.

\begin{figure}
\centering\includegraphics[width=8cm]{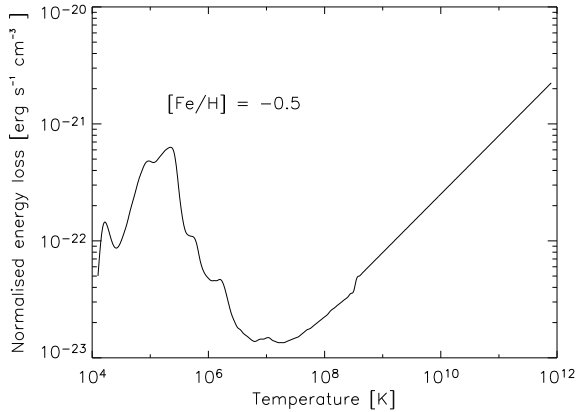}
\caption{Cooling function as a function of plasma
temperature. Note that the effects of line cooling at lower
temperatures mean that the gas cools much more rapidly at lower
temperatures than if we just assume bremsstrahlung. At high
temperatures the cooling is dominated by bremsstrahlung.}
\label{fig:coolf}
\end{figure}

In order to properly take into account the effect of chemical
enrichment and metallicity on the radiative cooling we adopt the
cooling function calculated by \cite{sutherland}. The function is
suitable for low density astrophysical plasma and was obtained from
the study of plasmas cooling in both equilibrium and non-equilibrium
situations in the temperature range $ 10^{4}-10^{8.5} $ K. By using
this cooling function we are able to account for the effects of
emission-line cooling which become significant for temperatures below
$ \sim 10^{7} $ (see figure \ref{fig:coolf}).  We assume the
metallicity to be half of the solar value \citep{edge91} and that the
cooling function depends only on temperature and metallicity, but not
density.  For temperatures below $ 10^{4} $ K the cooling effect is
'switched off' meaning that we do not accurately model cooling below
this temperature. However, for none of our simulations were such low
gas temperatures reached.

In order to implement the plasma thermal conductivity correctly it is
essential to know whether the electron mean free path is less than the
scale length of the temperature gradient. The scale length of the
temperature gradient can be defined as $ l_{\rm T} \equiv T/\nabla T $
\citep{sarazin}. For electron mean free paths which are greater than
the scale length of the temperature gradient the thermal conduction is
said to 'saturate' and the heat flux approaches a limiting value
\citep{cowie77}. For electron mean free paths much less than the
temperature gradient the heat flux depends on the coefficient of
thermal conductivity and the temperature gradient.

The mean free path of an electron ($ \lambda_{\rm e} $) in an
unmagnetized plasma is given by \citep{sarazin}

\begin{equation} \label{eq:mfp}
\lambda_{\rm e} = \frac{3^{3/2} \big( k_{\rm B} T_{\rm e} \big)^{2}}{4
\pi^{1/2} n_{\rm e} e^{4} \ln \Lambda_{\rm c}}~[\rm cm]
\end{equation}

Substituting typical values for near the centre of the cluster ($
n_{\rm e} \sim 0.1 \rm cm^{-3}, T_{\rm e} \sim 10^{7} \rm K $) gives
an electron mean free path of roughly 10 pc where the length of the
temperature gradient is of the order of kiloparsecs. For the outer
cluster regions ($ n_{\rm e} \sim 10^{-4} \rm cm^{-3}, T_{\rm e} \sim
3 \times 10^{7} \rm K $) we find a mean free path of $ \sim $ 20 kpc
where the scale length of the temperature gradient is of the order of
hundreds of kpc. This means that we are able to use the usual,
'non-saturated' form for the thermal conductivity at all radii for the
Virgo cluster.

Using the above result, we implemented thermal conduction and
viscosity with the transport coefficients given in equations
(\ref{eq:therm}) and (\ref{eq:visc}), suppressed by the factor, $f$,
into the simulation software. We use the same suppression factor for
both conductivity and viscosity as the two processes are intimately
linked \citep[e.g.][]{instab05}. In addition we also include a control
run in which thermal conduction and viscosity are absent. For the
cases in which we consider the effects of suppressed thermal
conductivity and viscosity, the suppression factor is constant
throughout the ICM. In reality the suppression factor is unlikely to
be uniform due to varying magnetic field strength and tangling
length. We note that unless the central regions of galaxy clusters are
turbulent such that the thermal conductivity is suppressed only by a
factor of a few \citep{zakamska03}, then it is likely that the
tangling length of the magnetic field is shorter in the central
regions (due to compression). Therefore the thermal conductivity may
be smaller in the centre compared to the outer regions. The effect of
this would be to reduce thermal conduction where it is required most.

When taking into account thermal conductivity as a source of heat for
balancing radiative losses, i.e. $\epsilon _{\rm cond} \sim \epsilon
_{\rm rad}$, it is necessary to have sufficient energy in the outer
regions of the cluster to achieve this. We can consider the outer
regions of a cluster to be an infinite heat bath, however, larger
volumes take longer to simulate. Simulating smaller volumes is
quicker, but it may starve the simulated cluster of the energy that
thermal conduction requires. The size of the computational box was
fixed as a cube of dimension 648 kpc, with the cluster centre situated
at the box centre. For a computational box of this size the regions
outside 5 kpc (inside which the cooling catastrophe was observed to
occur in preliminary simulations) contain roughly 10,000 times more
energy than can be supplied to the central 5 kpc over the lifetime of
a simulation (roughly 2-5 Gyr).

It is also worth pointing out that a problem noted by \cite{dolag04}
in which simulations involving radiatve cooling and star formation
found an increase in temperature at the cluster centre was also
observed in our preliminary simulations. We attribute this effect to
limited numerical resolution, since this effect of compressional
heating was eliminated by increasing the number of computational
blocks in the central region of our cluster (see Appendix).

\subsection{The Simulations}\label{sims}

The nature of the problem we are considering here is 1-dimensional in
many respects; this includes the initial conditions and the
non-evolving gravitational potential. In general the physical
mechanisms operating in this setup will result in mainly radial flow
of the gas. It is possible to achieve much higher spatial resolution
in a 1-d compared to a 3-d simulation. However, by employing a 3-d
geometry we allow for any non-radial processes which may occur, for
example: the growth of non-spherical modes of the thermal instability
or non-radial mass flow which is likely to occur near the cluster
centre.

In order to determine any differences between employing 3-d or 1-d
coordinate systems we have performed 1-d spherically symmetric
simulations for two cases: i) the absence of thermal conduction and
viscosity and ii) the limit of full Spitzer thermal conductivity and
viscosity. The results of these simulations are similar to those
performed using a 3-d coordinate system, and are given more completely
in Appendix \ref{app:1d3d}. The main observable difference is that the
evolution of 1-d spherically symmetric clusters is slightly more rapid
than in the 3-d case. We consider this behaviour consistent with the
inherent differences between the hydrodynamic equations in 1-d and
3-d. For example, there are extra dissipation terms both including and
excluding the viscosity in the 3-d case which are absent in the 1-d
case (see Appendix \ref{app:1d3d}) thus any non-spherically symmetric
processes will never be allowed to develop in 1-d. We therefore
proceed with 3-d simulations since we believe that the 3-d effects are
important in the overall evolution of the gas.

We performed 4 simulations to which we assign labels such that the
number preceeding $ \kappa $ is the suppression factor, $f$.

\begin{enumerate}
\item radiative cooling alone (zero thermal conduction and viscosity)
(0$ \kappa  $).
\item radiative and thermal conductivity at one
hundredth of the full Spitzer value (0.01$ \kappa $).
\item radiative and thermal conductivity at one
tenth of the full Spitzer value (0.1$ \kappa $).
\item radiative and thermal conductivity at the full
Spitzer value (1$ \kappa $).
\end{enumerate}

Table 1 provides a summary of the simulations.  

Each simulation was stopped after the onset of a cooling catastrophe
in the cluster centre. As a criterion for a cooling catastrophe we
used the timestep adopted by the simulation software. If this timestep
became insignificant compared to the duration of the entire
simulation, we stopped the simulation. This occured for central
temperatures of less than $ \sim 10^{7} $ K in all simulations.

\begin{table*} \label{tab:1}
\centering
\begin{minipage}{140mm}
\begin{tabular}{rlc}
\hline
 Name  & Supression factor  & Simulated time (yrs) \\
\hline
  0$\kappa $ & 0 & 1.29 $\times 10^{9} $ \\
\hline
  0.01$\kappa $  &  1/100 & 1.35 $\times 10^{9} $ \\
\hline
  0.1$\kappa $ &  1/10 & 1.89 $\times 10^{9} $ \\
\hline 
  1$\kappa $ & 1 &  4.72 $\times 10^{9} $ \\
\hline
\end{tabular}
\caption{Summary of the five main simulations}
\end{minipage}
\end{table*}

\section{Results}

\subsection{Temperature and Electron Number Density Profiles} \label{res1}

\begin{figure*} 
\centering \includegraphics[width=14cm]{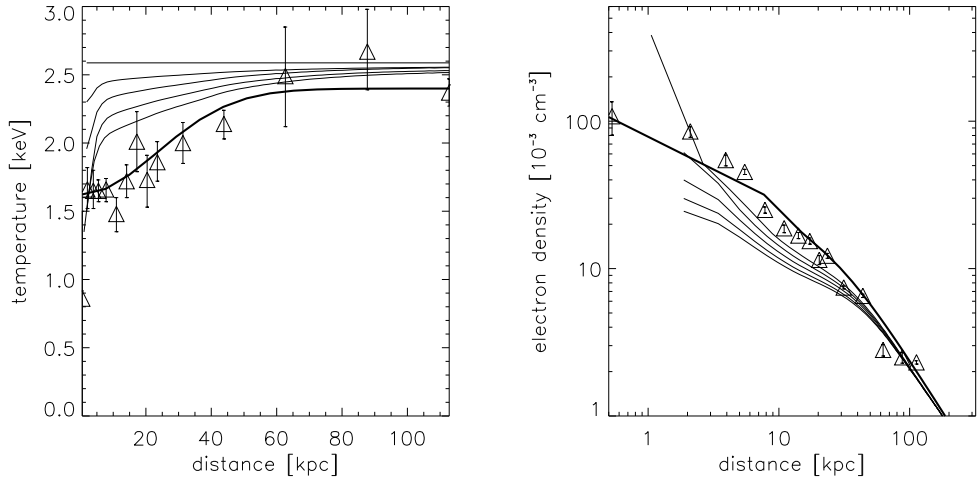}
\caption{Temperature and density profiles evolving with
time for simulation 0$ \kappa $ (see table 1). The thick lines for
both temperature and density are the functions fitted to the data
points (triangles) by Ghizzardi et al. (2004).  The top line in the
temperature plot shows the temperature profile after $3.17 \times
10^{8} $ yr and the bottom line at time of the end of the
simulation. These times are given in table 1 for each simulation. The
intermediate lines represent the temperatures at intervals of $ 3.17
\times 10^{8} $ yr after the top temperature profile. The temporal
sequence of the lines is reversed (bottom to top) in the density plot.}
\label{fig:0sp}
\end{figure*}

\begin{figure*}
\centering \includegraphics[width=14cm]{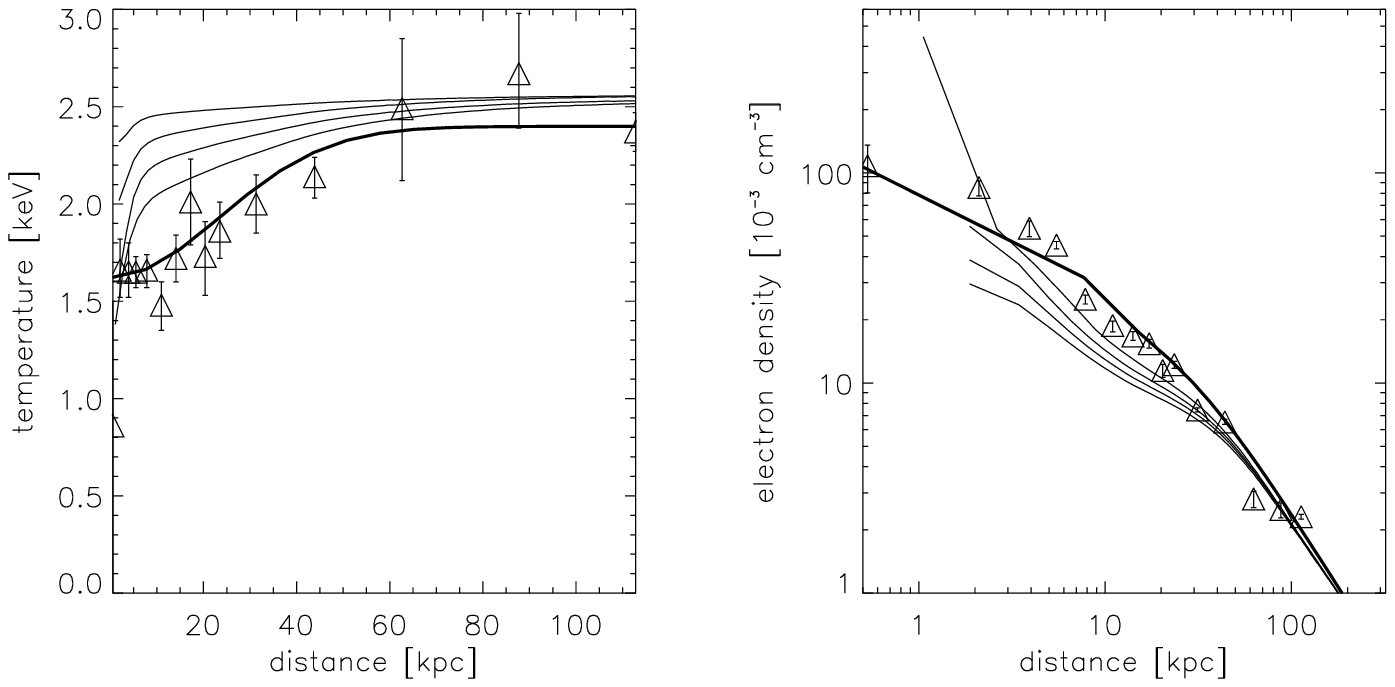}
\caption{\label{fig:100sp}Temperature and density profiles evolving
with time for simulation 0.01$ \kappa $.}
\end{figure*}

\begin{figure*} 
\centering \includegraphics[width=14cm]{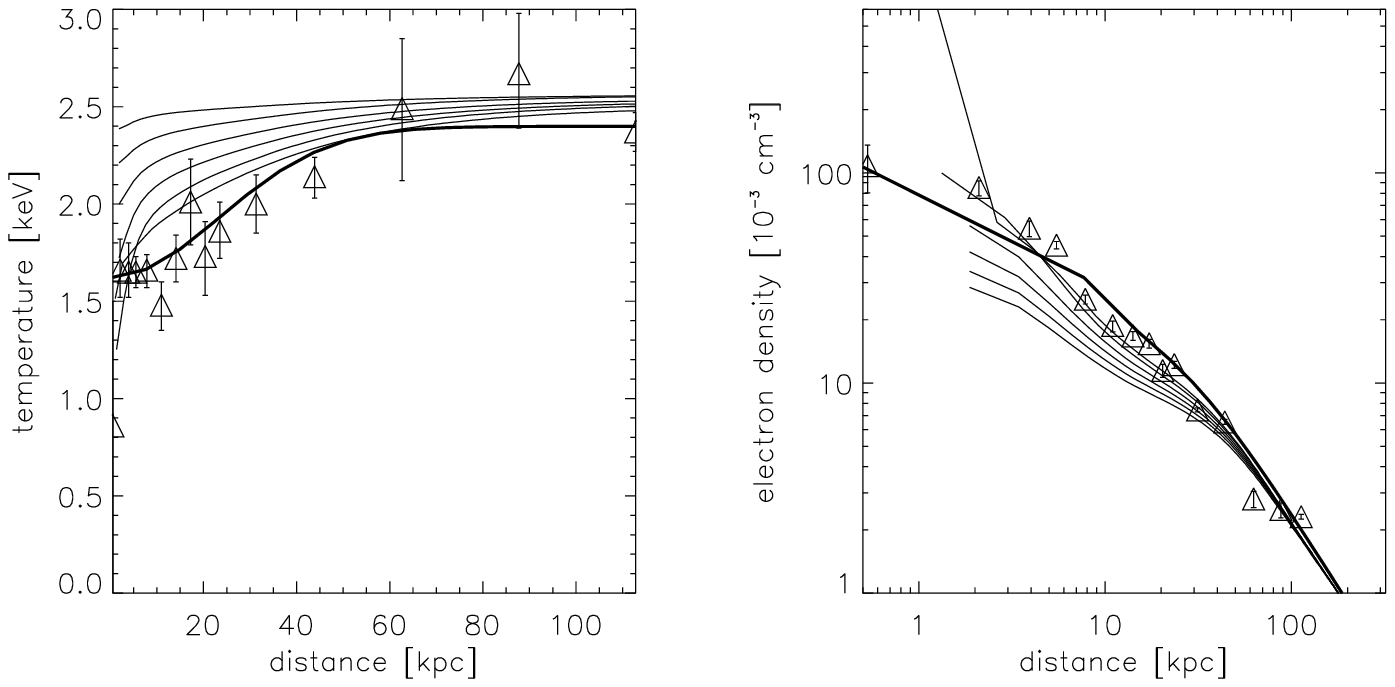}
\caption{\label{fig:10sp}Temperature and density profiles evolving
with time for simulation 0.1$\kappa $.}
\end{figure*}

\begin{figure*}
\centering \includegraphics[width=14cm]{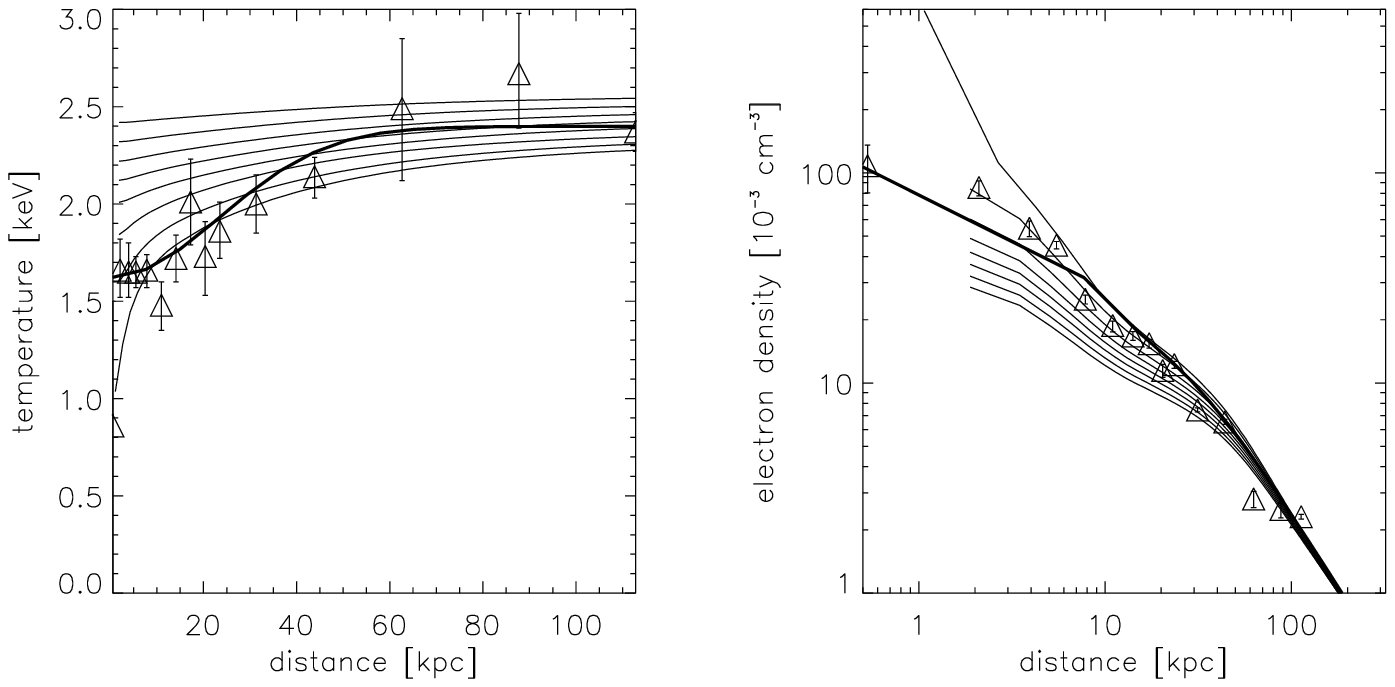}
\caption{\label{fig:sp} Temperature and density profiles evolving with
time for simulation 1$ \kappa $. Note that in this figure the
temperature and density are plotted at intervals of $6.34 \times
10^{8} $ yrs. }
\end{figure*}

We present spherically averaged profiles, from our simulations, for
both temperature and density, rather than 1-d slices through the
cluster. This is done by defining a number of spherical concentric
shells at a number of radii, in our case 200, from the cluster centre
and averaging the temperature and density between adjacent shells. The
temperatures and densities are compared to observations in figures
\ref{fig:0sp} to \ref{fig:sp}.

The temperature and electron number density profiles are plotted at
intervals of $ 3.17 \times 10^{8} $ yrs for all, except simulation 1$
\kappa $ which is plotted for every $ 6.34 \times 10^{8} $ yrs, and
for the final time of the simulation (see table 1). We also include
the commen initial conditions for simulation 0$\kappa$ only. For each
simulation we note that the central density increases at all times
during the ICM evolution. However, the behaviour of the gas
temperature is rather more complex: during the early phases of
evolution of the ICM the temperature falls, but appears to rise after
the onset of the cooling catastrophe (see also section \ref{res4}).

The cooling catastrophe is characterised by the relatively sudden and
dramatic increase in the central density over a short period of time
(few $\times 10^{8}$ yrs). It should be noted that no physical value
of thermal conductivity can prevent the occurance of a cooling
catastrophe.

One result from our simulations is that regardless of whether thermal
conductivity and viscosity are present, both the temperature and
density profiles eventually approach generic profiles. After 2 $\times
10^{9}$ yrs even the simulation with full Spitzer thermal conductivity
develops into a cooling catastrophe with temperature and density
profiles similar to the other simulations.

The temperature profiles are characterised by a narrow, but deep
central dip which deepens with time such that the central temperatures
fall below the observed approximate minimum of $ \rm T_{\rm vir}/3 $
\citep[e.g.][]{allen01}. Qualitatively, the main effect of thermal
conduction is an increase of the width of the central temperature
dip. This is due to thermal conduction tranporting energy down the
temperature gradient towards the cluster centre. This reduces the rate
at which the temperature falls in the cluster centre, but also
increases the effective rate at which the gas cools at larger
radii. The time taken to evolve into this profile is roughly inversely
proportional to the magnitude of the thermal conductivity.

The broader temperature dip associated with larger thermal
conductivities is accompanied by larger densities in the same region
of the cluster. This has the same physical explanation as above. Due
to thermal conductivity, the gas at larger radii loses more energy
than it would otherwise do due to radiative losses. This prompts the
inflow of gas at larger radii creating a build-up of material in a
larger region of the cluster centre than for lower values of the
thermal conductivity.

The density profiles can be adequately described by a strongly
centrally peaked distribution in which the peak grows with time. As
energy is radiated away most rapidly in the densest regions, the
pressure support is lost from these central regions and the weight of
the overlying layers compresses the central gas thus increasing the
density dramatically. The fall in central temperature and rise in
central density that we observe are both the obvious signs of a strong
cooling flow and eventually a cooling catastrophe in which large
quantities of gas are being deposited into the central regions of the
cluster.

It is evident from figures \ref{fig:0sp} to \ref{fig:sp} that for
increasing thermal conductivity the fit of the simulated density
profiles to the observational data improves at larger radii. However,
the profiles are not stable over sufficiently long timescales,
e.g. the Hubble time. Thus, thermal conduction only delays the
occurance of a cooling catastrophe and cannot simulataneously
reproduce temperature and density profiles that are consistent with
observations.

\subsection{Emissivity Profiles} \label{res2}

From the simulated density and temperature data it is possible to
derive X-ray surface brightness maps for our radiative cooling
function. Previously we have compared the density and temperature
distributions of our simulated clusters with those derived from
observations of the X-ray surface brightness of the gas, since these
are the fundmental variables used by FLASH. However, calculating the
surface brightness for our simulations allows for a more direct
comparison with the observational data. In practice the surface
brightness along a given line of sight is dominated by emission from
gas at the smallest radii from the cluster centre along that
particular line of sight. Thus we calculate the emissivity of the gas
($\epsilon $), i.e. the rate of energy loss per unit volume, and plot
this as a function of radius. The bolometric emissivity is given by

\begin{equation} \label{eq:surfbright}
\epsilon = n_{e} n_{p} \Lambda_{\rm rad}(T,Z) ~[{\rm erg
~s^{-1}cm^{-3}}]
\end{equation}

\begin{figure*}
\begin{minipage}[b]{.4\linewidth}
\centering\includegraphics[width=\linewidth]{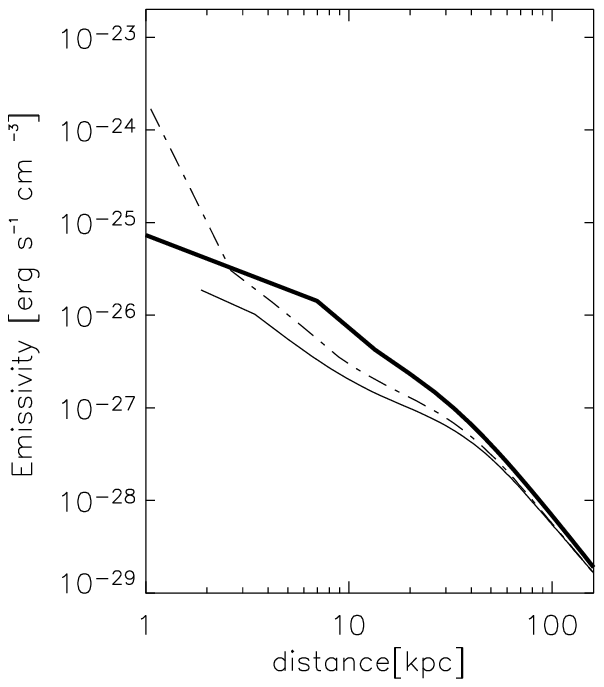}
\end{minipage}\hfill
\begin{minipage}[b]{.4\linewidth}
\centering\includegraphics[width=\linewidth]{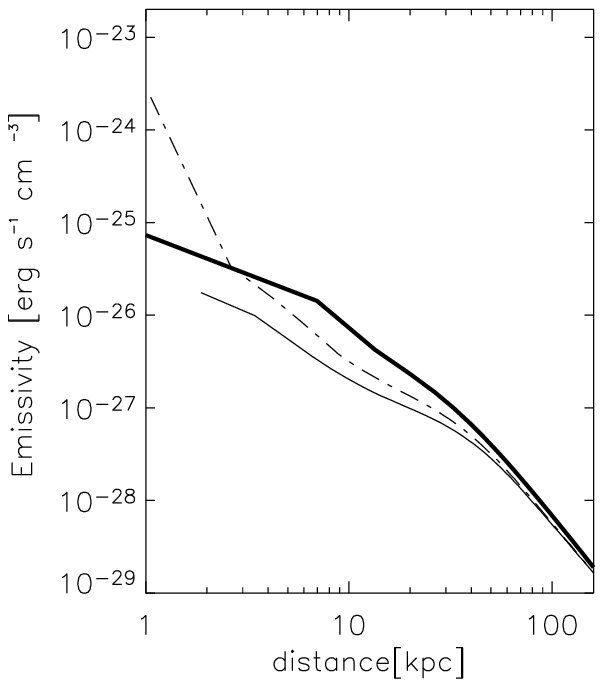}
\end{minipage}\hfill

\caption{Emissivity profiles for (from left to right) simulation 0$
\kappa $ and 100$ \kappa $. The thick line shows the observed
emissivity given by Ghizzardi et al. (2004), the thin solid line shows
the emissivity after $6.34 \times 10^{8}$ yrs; the thin dashed line
shows the emssivity at the end point of each simulation as given by
table 1.}
\label{fig:xrb1}
\end{figure*}

\begin{figure*}
\begin{minipage}[b]{.4\linewidth}
\centering\includegraphics[width=\linewidth]{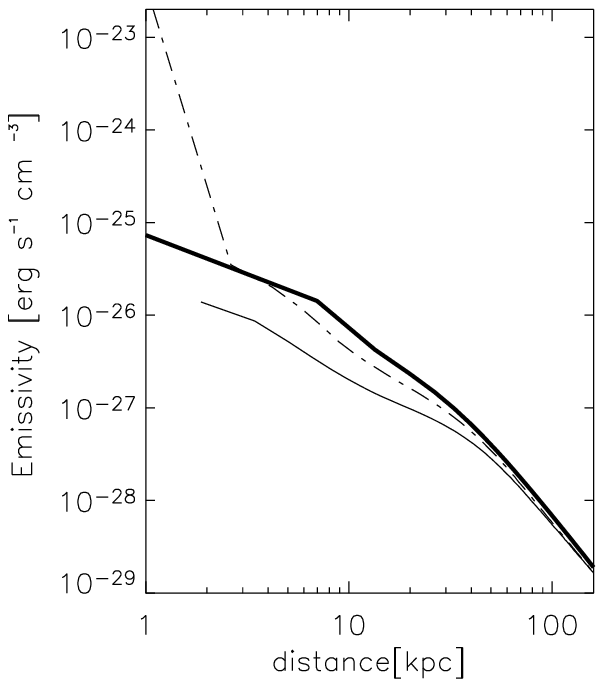}
\end{minipage}\hfill
\begin{minipage}[b]{.4\linewidth}
\centering\includegraphics[width=\linewidth]{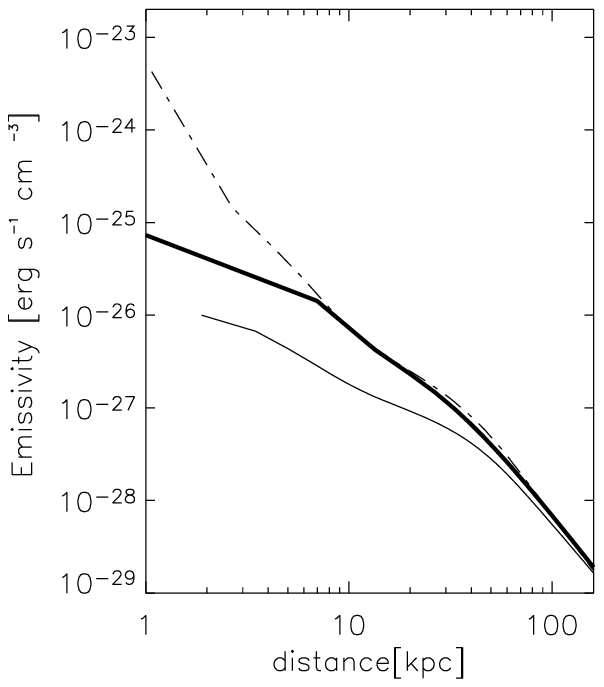}
\end{minipage}
\caption{Emissivity profiles for (from left to right) simulation 0.1$
\kappa $ and 1$ \kappa $. The definitions for each line are as in
figure \ref{fig:xrb1}.}
\label{fig:xrb2}
\end{figure*}

Figures \ref{fig:xrb1} to \ref{fig:xrb2} show the emissivity after
$6.34 \times 10^{8}$ yrs and at the final point in the simulation
compared to the currently observed emissivity calculated from the work
of \cite{ghizzardi04}.

XMM-Newton observations revealed that the X-ray surface brightness was
less centrally peaked than expected \cite[e.g.][]{missing02}. From
figures \ref{fig:xrb1} to \ref{fig:xrb2} it is evident that our
simulated clusters always develop a strong central peak in the surface
brightness distribution, whether or not thermal conduction is taken
into account. Such a peak is to be expected given the density profiles
in section \ref{res1}, but is not present in the observational
data. This again indicates that an extra heating mechanism in the
central few kiloparsecs is required to prevent the excessive central
cooling.

We note that the general fit at large radii to the observational data
seems to improve as the value of the thermal conductivity is
increased. In addition, the emissivity profiles show more excess
emission out to larger radii for the simulations with a larger thermal
conductivity. Thus, the emissivity plots re-iterate the previous
observation from the density profiles that for higher thermal
conductivities there is more mass in the central 20-50 kpc than for
lower values of thermal conductivity. This is consistent with the mass
flow rates which are larger at all radii for later times in the
presence of high thermal conductivity than those for lower thermal
conductivity (see section \ref{res5}). Hence, in order to better match
the observations by preventing a cooling catastrophe it appears that
the presence of a larger thermal conductivity actually requires the
input of more thermal energy. In addition, this heat source would need
to distribute its energy out to larger radii for the cases of larger
thermal conductvity. The extra energy input required is roughly
proportional to the extra volume which requires heating, and so an ICM
with large thermal conductivity could require 100-1000 times more
energy than for smaller thermal conductivities.

\subsection{Effective Adiabatic Index} \label{res3}

As an additional diagnostic for the behaviour of the cluster gas we
calculate the effective adiabatic index defined by $\gamma_{\rm eff}
\equiv {\rm d} \ln P/ {\rm d} \ln \rho $ where $ P $ is the gas
pressure, and $ \rho $ is the gas density.  Following the method of
\cite{birnboim03} we calculate $\gamma_{\rm eff} $ for a situation in
which the volume of gas is constant, but is subjected to heating and
cooling.  The pressure in terms of the internal energy per unit mass,
$e$, is given by

\begin{equation} \label{eq:pressure}
P = (\gamma -1)e \rho~[\rm{erg~ cm^{-3}}],
\end{equation}
where $\gamma $ is the true adiabatic index (e.g. 5/3 for a monatomic
gas)

The rate of change of gas pressure with time is given by
\begin{equation} \label{eq:pdot}
\dot{P} = (\gamma -1)(\dot{e}\rho +e\dot{\rho})~[\rm{erg~ cm^{-3}~
s^{-1}}].
\end{equation}

The rate of change of internal energy per unit mass for a constant
volume of gas subjected to heating and cooling is given by

\begin{equation} \label{eq:edothq}
\dot{e} = h - q~[\rm{erg~ g^{-1}}],
\end{equation}
where h and q are the heating and cooling functions per unit mass
respectively. 

By converting all the relevant quantities to be compatible with those
we have defined previously and using the definition of the effective
adiabatic index we find $\gamma_{\rm eff}$ to be given by

\begin{equation} \label{eq:gameff}
\gamma_{\rm eff}= 1 + (\gamma -1)\bigg(\frac{n^{2}H -
 n^{2}\Lambda_{\rm rad}}{\dot{n} k_{\rm B}T}\bigg),
\end{equation}
where H is the heating function (the heating analogue of the cooling
function, $\Lambda_{\rm rad}$), $\dot{n}$ is the rate of change of
number density with time and the other symbols have their usual
definitions.

In this derivation the nature of the heating and cooling mechanisms is
not important, however, we choose to represent them in terms of
quantities that we have already defined. From equation
(\ref{eq:gameff}) we see that in the limit where either there is no
heating or cooling or these competing effects balance, the effective
index is 1, which corresponds to an isothermal equation of
state. Furthermore, we note that in the cases where no heating occurs,
$\gamma_{\rm eff}$ is always less than one and for extreme cases will
tend to zero as $ \dot{n} $ tends to $ n/t_{\rm cool}$ (where $t_{\rm
cool}$ is the radiative cooling time). For a constant pressure cooling
flow $\gamma_{\rm eff}$ will be equal to 1/3. In the cases where
heating occurs this will have the effect of increasing $\gamma_{\rm
eff}$. Using this equation we predict for the simulations which
include thermal conductivity at a significant fraction of the Spitzer
value that the gas exhibits a $\gamma_{\rm eff}$ that is greater than
for lower values of thermal conductivity, at least in the cluster
centre where radiative cooling is strongest.

\begin{figure}
\centering \includegraphics[width=7cm]{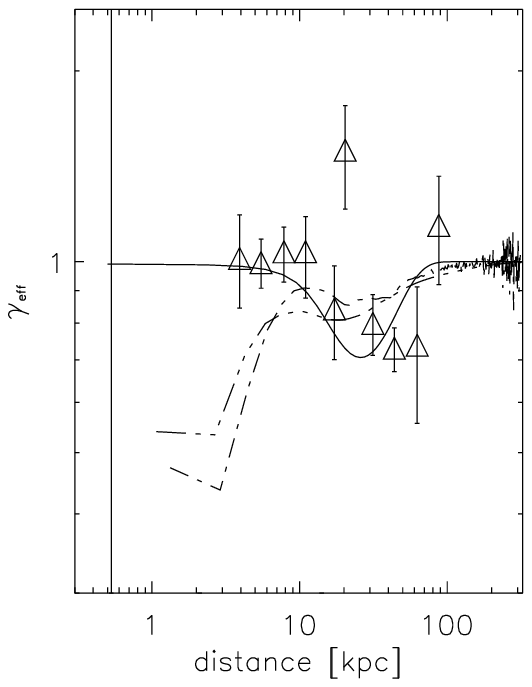}
\caption{Comparison of the effective adiabatic index for simulation
1$\kappa $ (triple dot dash), 0$\kappa$ (dot dash) and the effective
index from the fits to the observational data (solid line) and the
data points. The triangles show the observational data that the fits
are based on. At no point in time does the simulation data, within the
central 15 kpc, approximate to the observational data. Therefore, for
clarity, the $\gamma_{\rm eff}$ profiles for the simulation data are
calculated for times where the central temperature is near its lowest
point. The vertical line near the left of the figure is the error bar
from the data point closest to the cluster centre.}
\label{fig:gameff}
\end{figure}

In figure \ref{fig:gameff} we compare the effective adiabatic index
obtained from the results of \cite{ghizzardi04} with $ \gamma_{\rm
eff} $ for simulations 0$\kappa$ and 1$\kappa$. We note that in the
absence of thermal conductivity $\gamma_{\rm eff}$ is roughly equal to
1 (isothermal) in our simulation only in the outer regions of the
cluster, but falls steeply inside the central 10 kpc towards the
central value given in table 2. The upturn at roughly 3 kpc is
probably due to compressional heating towards the end of the simulated
time. We also note that there is a minimum at roughly 25 kpc and a
peak at roughly 15 kpc. For the full Spitzer value of thermal
conductivity, $\gamma_{\rm eff}$ follows roughly a similar pattern to
the above except that the variation is smoother. In addition, the
central effective adiabatic index is higher due to the heating effect
of thermal conductivity. In comparison, it is evident that for the
functions fitted to the observational data by \cite{ghizzardi04}
$\gamma_{\rm eff}$ is isothermal throughout the cluster except for
around 30 kpc where we observe a minimum similar to that noted in the
simulation data. Of particular interest is the fact that there are 4
data points between the radii of 4-12 kpc with an effective adiabatic
index of roughly 1 that are not consistent with either of the two
simulations. By equation (\ref{eq:gameff}) this value of the observed
$\gamma_{\rm eff}$ must be due to heating balancing cooling. From the
simulations we can see that such a profile for $\gamma_{\rm eff}$ is
unattainable with thermal conductivity alone. Therefore, this is
strong evidence that the Virgo cluster is being heated in the central
regions by a mechanism other than thermal conduction.

Despite the evidence for heating we are unable to make any strict
quantitative predictions about the level of thermal conduction in
Virgo. However, according to the observational data $\gamma_{\rm eff}$
appears to vary significantly throughout the cluster. Our results
suggest that large values of the thermal conductvity tend to smooth
out variations of $\gamma_{\rm eff}$. In addition, since thermal
conductivity acts to reduce the temperature gradient, any heating
source located near the centre of a cluster in which thermal
conduction operates at near the Spitzer value is likely to result in a
much flatter temperature profile than is currently observed. In the
case where thermal conduction is heavily suppressed, it is possible
that an additional heat source could maintain the temperature in the
cluster centre at the observed level while allowing the region outside
to continue to cool radiatively and thus produce the observed wide
bowl shape. Thus although we are unable to put strict constraints on
the value of thermal conductivity in Virgo, our results suggest that
it is probably suppressed by a factor of at least 10 or more. Recent
work by \cite{brueggen03} and \cite{ruszbegel02} suggests that, in 1-d
at least, it is possible to achieve a steady state between radiative
cooling and simultaneous heating from an AGN and thermal conduction
suppressed by factor similar to our suggestion.


\subsection{Central Temperature and density} \label{res4}

\begin{figure*}
\centering \includegraphics[width=7cm]{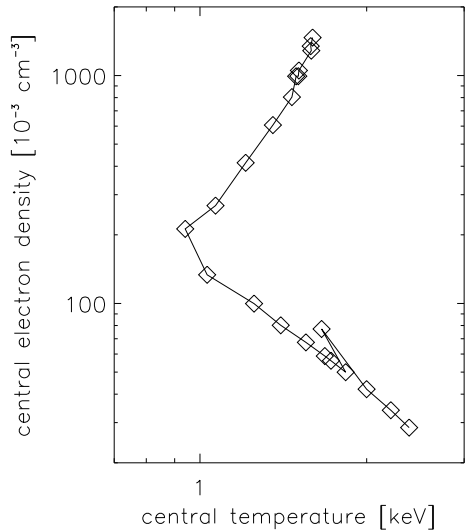}
\caption{Central density against central
temperature evolving as a function of time for simulation
0.01$\kappa$. The bottom right of the plot corresponds to the initial
conditions and the system progresses upwards with passing time.}
\label{fig:tminnmax}
\end{figure*}

It is also instructive to plot the central density against central
temperature, at various points in time throughout each
simulation. Figure \ref{fig:tminnmax} shows this for simulation
0.01$\kappa$. Each simulation shows the same three main
characteristics: an initial power-law dependence between falling
temperature and increasing density followed by a turning point near
which the temperature begins to increase up to an asymptotic value
while the density continues to increase. The results for each
individual simulation are given in table 2 where $a$ and $b$ are
defined by a power-law relating temperature and density: $ n = b
T^{a}$. Given that the early evolutionary behaviour of the central
temperature and density is accurately described by a power-law it is
also appropriate to describe this behaviour in terms of an effective
adiabatic index, $\gamma_{\rm eff}$ which for the specified power-law
is given by $1 +1/a$. The results given in table 2 show that the
intial cooling phases are well described by power-laws and that $ a $
decreases for increasing values of thermal conductivity and $ b $,
which relates to the entropy, increases. $\gamma_{\rm eff}$ increases
for larger thermal conductivities because of the heating effect of
thermal conduction in the cluster centre.

After the initial cooling in which the temperature falls with time, we
observe without exception a period in which the central temperature
begins to rise to an asymptotic value of roughly 1.6 keV which appears
to be the same for each simulation. We have been unable to investigate
the behaviour past this asymptotic point due to the smallness of the
timestep at the time when this behaviour was observed.

The temperature rise is probably due to the cooling catastrophe which
occurs after $10^{9}$ yrs. The rapid inflow of material in the cluster
centre results in some compressional heating. After a certain time the
central temperature has reached its maximum and begins to decrease
again. Throughout this behaviour the density continues to increase
while the temperature probably oscillates around the asymptotic value
to some degree.

It is possible that the central collapse causes some reheating of the
very centre of the cluster. Note that we did not observe this 'bounce'
for the trial simulations conducted for 1/8th of the volume, but with
the same spatial resolution in the gas flow, of the simulations
presented here. We also find no evidence for a shock occuring near or
after the turning point. It appears that the power-law observed in
these simulations and the other notable characteristics are unaffected
by increasing the numerical resolution, see Appendix \ref{app:numres}.

\begin{table*}
\centering
\begin{minipage}{120mm}
\caption{Summary of the parameters describing the behaviour of the
central temperature and density for each simulation. The temperatures
are given in keV and densities in units of $10^{3} \rm {cm}^{-3}$}
\begin{tabular}{cccccc} \hline
$\kappa/\kappa_{\rm S}$ & $a$ & $b$ & $\gamma_{\rm eff}$ & turning point(T,n) & asymptote(T)\\ 
\hline 
0 & -1.60$\pm$0.025 & -30.7$\pm$0.41 & 0.375$\pm$0.006& $\simeq$0.84, $\simeq$151 & $\simeq$1.6\\ 
\hline
 0.01 &-1.67$\pm$0.016 &-29.7$\pm$0.27 & 0.40$\pm$0.004& $\simeq$0.89, $\simeq$149 &$\simeq$1.6\\
 \hline 
0.1& -1.96$\pm$0.119 & -24.5$\pm$1.95 & 0.49$\pm$0.029& $\simeq$0.94, $\simeq$212 & $\simeq$1.6\\ 
\hline 
1 & -2.32$\pm$0.070 &-18.4$\pm$1.17 & 0.57$\pm$0.017& $\simeq$0.82, $\simeq$270 & $\simeq$1.6\\
 \hline
\end{tabular}
\end{minipage}
\end{table*}


\subsection{Mass deposition rates} \label{res5}

In this Section we calculate mass flow rates as a function of
radius. Thus we can directly estimate the rate at which gas flows in
to the cluster centre for different thermal conductivities. In
addition, we also calculate theoretical mass flow rates and compare
them to the results from our simulations.

The mass deposition rate is given by the mass flux through a spherical
surface

\begin{equation} \label{eq:massdepos}
\frac{{\rm d}M}{{\rm d}t} = \int \rho \bm{v}.{\rm d}\bm{A} 
= 4 \pi r^{2} \rho v_{\rm r} ~[{\rm g~s^{-1}}],
\end{equation}
where $v_{\rm r}$ is the projection of the velocity vector on to the
radial vector.

\begin{figure*}
\begin{minipage}[b]{.4\linewidth}
\centering\includegraphics[width=\linewidth]{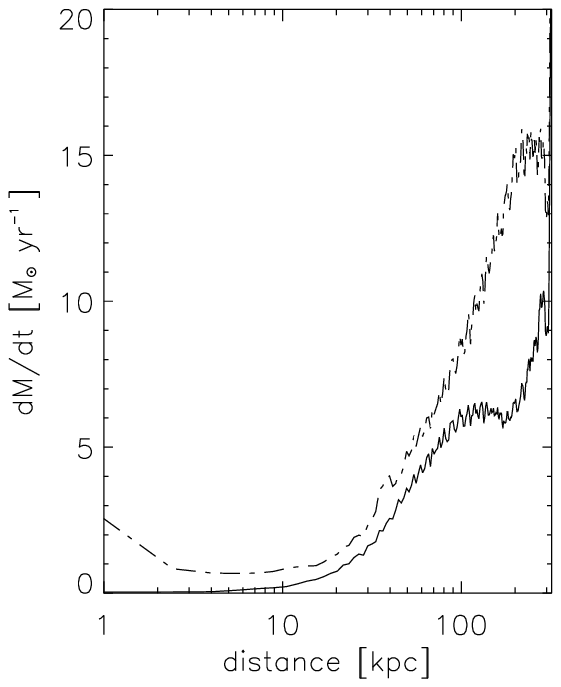}
\end{minipage}\hfill
\begin{minipage}[b]{.4\linewidth}
\centering\includegraphics[width=\linewidth]{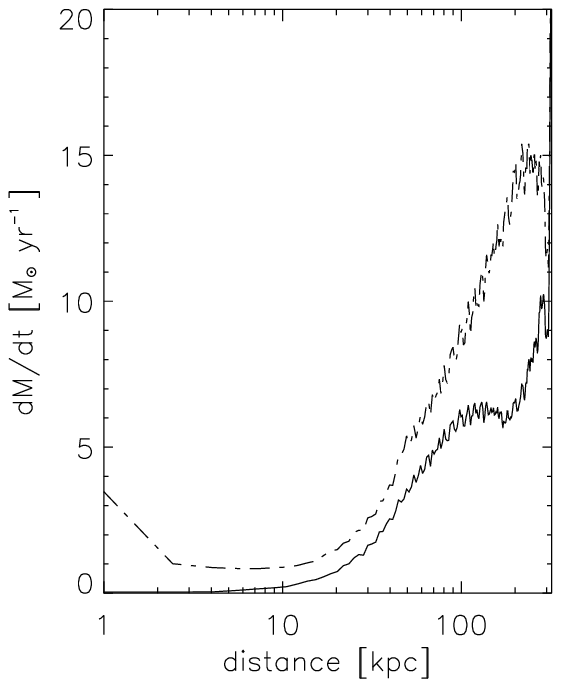}
\end{minipage}\hfill
\caption{Mass deposition profiles for (from left to right) simulations
0$ \kappa $ and 0.01$ \kappa $. In all plots the solid line shows the
mass flow rate at $ 6.34 \times 10^{8} $ yrs; the dashed line refers
to the end point in the simulation. The times of the end points are
given in table 1.}
\label{fig:mdot1}
\end{figure*}

\begin{figure*}
\begin{minipage}[b]{.4\linewidth}
\centering\includegraphics[width=\linewidth]{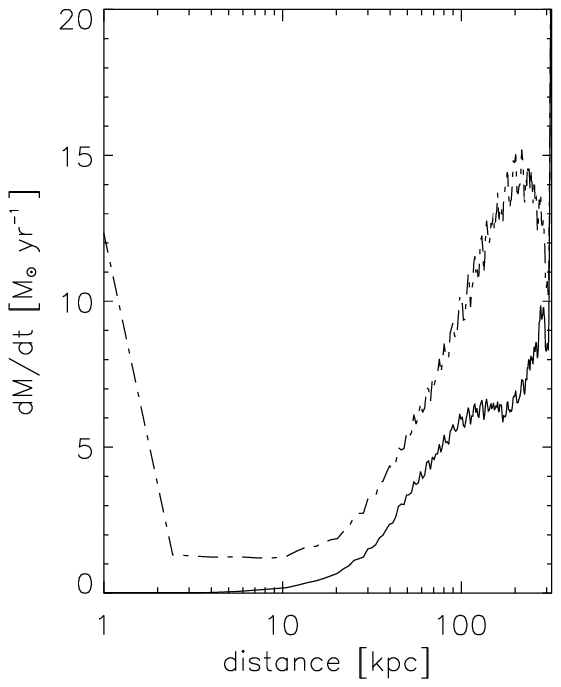}
\end{minipage}\hfill
\begin{minipage}[b]{.4\linewidth}
\centering\includegraphics[width=\linewidth]{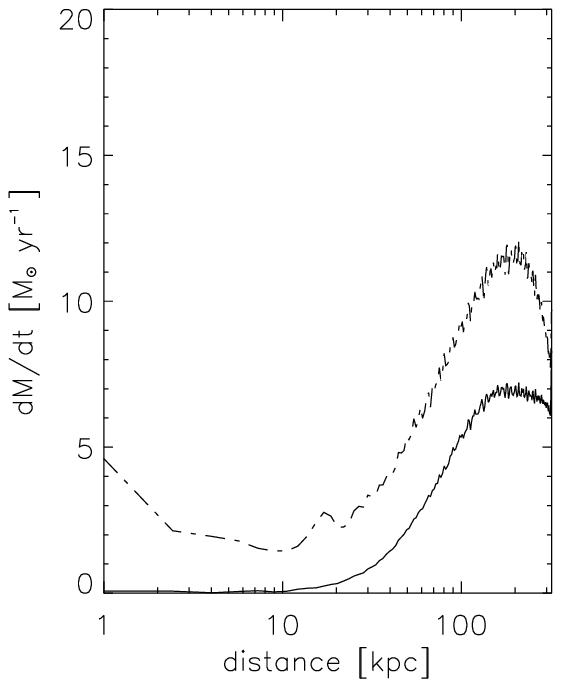}
\end{minipage}
\caption{Mass deposition profiles for (from left to right) simulations
  0.1$ \kappa $ and 1$ \kappa $. The dashed and solid lines are
  defined as in figure \ref{fig:mdot1}.}
\label{fig:mdot2}
\end{figure*}

Figures \ref{fig:mdot1} to \ref{fig:mdot2} show the mass deposition
rates as a function of radius for two different times. These figures
show the local mass flow rates and not the integrated mass flow rate,
M($<$r). The first line (solid) is the mass deposition rate profile
after $ 6.34 \times 10^{8} $ yrs, the second line (dashed) is the
profile for the final point in the simulation given in table 1.

For the mass flow rate profiles after $ 6.34 \times 10^{8} $ yrs the
flow rate tends to zero in the cluster centre in all
simulations. There is also a point near 150 kpc at which the flow rate
reaches a maximum. From here on we refer to this point as the
break-radius \citep{edgerb92,allen201}. The increase outside the
break-radius (at roughly 200 kpc) is probably not realistic, but is
due to edge effects in which the initial velocities are large, but
dissipate over the duration of the simulation.

For the later profiles, when the cooling flow is well established, the
central mass deposition rate increases up to several solar masses per
year in all cases. We also note that the break-radius has moved
further away from the cluster centre. The peak mass flow, at the break
radius, is smaller for larger values of thermal conductivity. This is
possibly a consequence of the action of shear viscosity which slows
the inflow of the gas. In addition, the width of the mass flow rate
peak increases with increasing thermal conductivity and viscosity,
indicating that in such cases there is more mass flowing towards the
centre and that the inflow velocity falls off less quickly with
radius. Thus it is possible that the central galaxies in clusters,
which are thermally conducting, may be more massive, by a factor of a
few, than those in cluster in which thermal conduction does not occur.

The calculated mass deposition rates are important for comparison with
observational results. \cite{edge01} finds the mass of molcular gas
around M87 to be less than $ 1.3 \times 10^{8}\rm M_{\odot} $. If the
Virgo cluster has been cooling for roughly 10 Gyr then the average
mass deposition rate over this period cannot be greater than 0.013
$\rm M_{\odot} $ yr$^{-1}$. We find that for a thermal conductivity at
the Spitzer value the mass deposition rate in the centre of the
cluster is negligible for the first $ \sim 10^{9} $ years, but rises
to several $\rm M_{\odot} $ yr$^{-1}$ thereafter. In addition, the
rate of mass deposition will continue to increase after this time as
the cooling catastrophe accelerates. For the cases with suppressed
thermal conductivity the central mass deposition rate is significant
from earlier times meaning that more gas is deposited in the cluster
centre for these simulations. Therefore, although the central mass
flow rates of a few $ \rm M_{\odot} yr^{-1}$ appear at first rather
modest, it is worth noting that the mass flow rates in the centre will
quickly exceed the observations` upper limit.

It is possible to compare the calculated mass flow rates from our
simulations with analytical predictions. In order to do this we
require both the density and velocity profiles. However, it is not
possible to solve the full hydrodynamic equations analytically for the
velocity profile as a function of time due to the complexity of the
problem. Instead we use both observational and simulation data to
constrain the functional form of the velocity profile at any given
time.

Figures \ref{fig:mdot1} and \ref{fig:mdot2} show that there is clearly
a break radius in the mass flow rate profiles which agrees with the
observations of \cite{edgerb92} and \cite{allen201}. However, unlike
these authors we plot the mass flow rate at each radius and not the
cumulative mass flow rate. Since the radiative cooling rate is
greatest near the cluster centre, gas in the central regions will lose
its pressure support. This means that not only would we expect the gas
infall velocity to tend asymptotically to zero for large radii, but
that the scale height of the velocity profile will increase as a
function of time as the less dense gas at larger radii has time to
cool. The radiative cooling time as a function of radius is given by
equation (\ref{eq:tcool}). We assume the cluster gas to be isothermal
and the density to follow a generalised $\beta $-profile. Rearranging
for the radius, we arrive at the cooling radius as a function of time

\begin{equation} \label{eq:rcool}
\frac{r_{\rm cool}}{r_{0}} = \bigg(\bigg( \frac{2 t \Lambda_{\rm rad0}
n_{0}}{3 k_{\rm B} T^{0.5}}\bigg)^{\frac{2}{3 \beta}} -1
\bigg)^{\frac{1}{2}},
\end{equation}
where $ r_{0} $ is the scale height of the density distribution. For
the Virgo cluster the value of $ \beta $ is roughly 0.47 for radii greater
than roughly 30 kpc.  

Equation (\ref{eq:rcool}) implies that after times of order 0.3 Gyrs
the velocity scale height grows proportional to $ t^{1/(3\beta)}$.

Further constraints on the velocity profile can be determined using
equation (\ref{eq:massdepos}) and what we already know about the gas
density of the Virgo cluster. The electron number density given by
equation (\ref{eq:virgdens}) implies that at large radii the density
scales as $ r^{1.4} $. Substituting this into equation
(\ref{eq:massdepos}) implies that in order for there to be a break
radius the infall velocity of the gas must fall off more quickly than
approximately $ r^{-0.6} $. Furthermore, by the observation that the
mass flow rate tends asymptotically to zero in the cluster centre
(before the cooling catastrophe occurs) it follows that, at least, in
the central region the velocity cannot fall off faster than $ r^{-2}
$. However, it should be noted that after the occurance of the cooling
catastrophe the central mass flow rate becomes finite. This suggests
that the velocity profile steepens at least in the central few
kiloparsecs and may be an indication that the gas is in free-fall
after the onset of the cooling catastrophe. Thus, for the Virgo
cluster, if the infall velocity follows a power-law ($ v \sim
r^{-\alpha} $) then $ \alpha $ may be constrained to lie between 0.6
and 2. In general, for any cluster in which the gas density may be
described by a $ \beta $-profile the constraints on $\alpha $ are (2 -
3 $\beta) \le \alpha \le $ 2.

In the absence of an analytical expression for the velocity profile we
adopt a velocity profile as $ v \sim \exp(-r^{2}/{r_{\rm s}}^{2}) $
which satifies the condition above, where $ r_{\rm s} $ is the scale
height of the velocity profile which is equivalent to the cooling
radius described in equation (\ref{eq:rcool}).

\begin{figure*}
\centering \includegraphics[width=14cm]{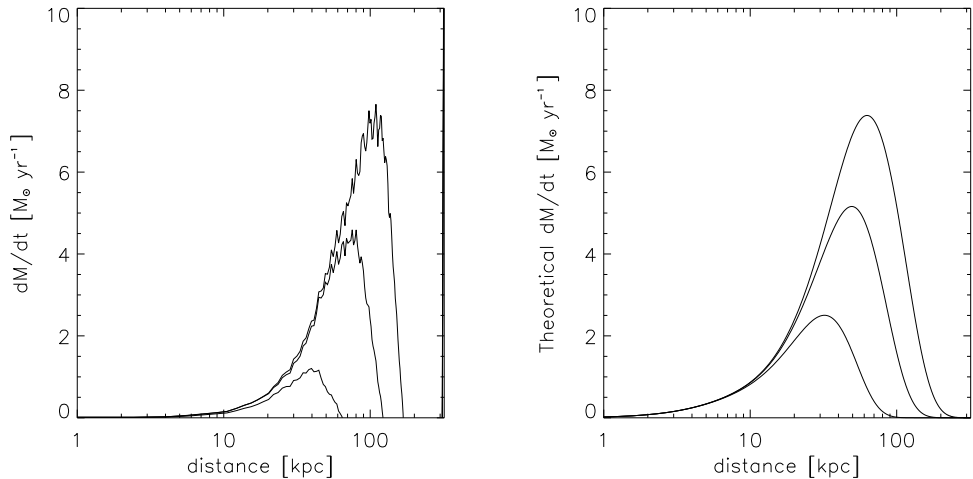}
\caption{Comparison of the mass flow rates for increasing time for
simulation 0.01$ \kappa $ data (left) and from simple theoretical
predictions at the same times.}
\label{fig:mdottheory100sp}
\end{figure*}

In order to understand the dependence of the mass flow rate on
velocity we use the density profile that we calculated for the initial
conditions of these simulations and the velocity profile described
above and vary only the velocity scale height with time. The peak mass
flow rate is determined by the velocity in the cluster centre, which
by comparison with the simulation we chose to be of order $ 10^{5}
{\rm cm~s^{-1}}$. The density profile can be assumed to remain fixed
since the density should not change significantly until a cooling
catastrophe occurs. However, a gradual increase in the central density
will inevitably occur due to the inflow of gas towards the cluster
centre. The results of this basic analysis are compared to the
simulation data in figure \ref{fig:mdottheory100sp}. It is obvious
that despite the simplistic nature of the analysis its predictions
match the simulation results reasonably well. In addition, the
analysis should hold for clusters in general and not just the Virgo
cluster. Thus our inflow velocity profile, $ v \sim
\exp(-r^{2}/{r_{\rm s}}^{2}) $, can be taken as a generic
approximation for the gas velocity in radiatively cooling clusters.

\section{Summary}

\label{conc}

Our results suggest that thermal conduction is unable to prevent the
occurance of a cooling catastrophe and can only increase the cooling
time in the cluster centre by a factor of a few. Tests using different
coordinate systems and spatial resolutions show that the developement
of a cooling catastrophe is inevitable. We do not consider this to be
a consequence of our chosen initial conditions since we allow the
cluster to evolve self-consistently, taking into account all relevant
physics of which we are aware, using standard intial conditions used
in cluster formation theory.

The temperature and density profiles from our simulated clusters fail
to simultaneously converge to the observational constraints. The mass
flow rates we calculate also suggest that significantly more mass is
deposited in the cluster centre than is observed. Somewhat contrary to
expectations we also find that the mass flow rates are also greatest
for the simulations in which thermal conductivion is most
important. The extra central concentration of mass this causes results
in emissivities with broader peaks than for lower thermal
conductivities. A consequence of this is that for these cases more
energy has to be injected, for example, by a central AGN than for more
weakly conducting cases to avert a cooling catastrophe.

These results are in agreement with \cite{voigt02} who found that in
their sample thermal conduction could only balance radiative losses in
the outer regions of cluster cores. There is a slight contradiction
with the findings of \cite{zakamska03} who found that the temperature
and density profiles of several galaxy clusters were consistent with a
steady state maintained by physically meaningful values of thermal
conduction. However, since \cite{zakamska03} do not take into account
the effect of line cooling, which is more significant at the lower
temperatures near cluster centres, and only look for the temperature
and density profiles which will ensure a steady state, it is unclear
whether this study and ours are comparable. Furthermore, neither of
\cite{voigt02} nor \cite{zakamska03} have made predictions about the
Virgo cluster. It would therefore be of interest to undertake a
similar study, to this, for a cluster for which \cite{zakamska03}
suggest thermal conduction could maintain a steady state.

The simulated and observed effective adiabatic index indicates that
there is a heat source at the centre of the Virgo cluster. In
addition, given this and the temperature and density profiles we
suggest that thermal conductivity is suppressed to at most 1/100 to
1/10 of the full Spitzer value. However, even if thermal conductivity
and viscosity are suppressed by a factor of 100, the thermal
conduction and viscous dissipation of sound waves generated in the ICM
could be significant.

For the overall energy budget we suggest that thermal conduction is
unlikely to be able to suppress the cooling in the majority of
clusters. The reason for this is simple: the physics underlying
thermal conduction and radiative cooling is too different for the two
processes to balance each other. The fact that these energy terms
cannot balance means that the temperature and density profiles are
likely to be unstable on cosmologically relevant timescales. Thermal
conductivity cannot be responsible for the universal temperature
profiles of clusters as observed by \cite{allen01}.

\section{Acknowledgements}

ECDP acknowledges support from the Southampton Regional eScience
centre in the form of a studentship. GP and CRK thank PPARC for
rolling grant support. The software used for this work in part
developed by the DOE-supported ASCI/Alliance Center for Astrophysical
Thermonuclear Flashes at the University of Chicago. We would like to
thank the anonymous referee for helpful comments.

\bibliography{database} \bibliographystyle{mn2e}

\begin{thebibliography}{}

\bibitem[\protect\citeauthoryear{{Allen}, {Fabian}, {Johnstone}, {Arnaud} \&
  {Nulsen}}{{Allen} et~al.}{2001}]{allen201}
{Allen} S.~W.,  {Fabian} A.~C.,  {Johnstone} R.~M.,  {Arnaud} K.~A.,
  {Nulsen} P.~E.~J.,  2001, MNRAS, 322, 589

\bibitem[\protect\citeauthoryear{{Allen}, {Schmidt} \& {Fabian}}{{Allen}
  et~al.}{2001}]{allen01}
{Allen} S.~W.,  {Schmidt} R.~W.,    {Fabian} A.~C.,  2001, MNRAS, 328, L37

\bibitem[\protect\citeauthoryear{{Birnboim} \& {Dekel}}{{Birnboim} \&
  {Dekel}}{2003}]{birnboim03}
{Birnboim} Y.,  {Dekel} A.,  2003, MNRAS, 345, 349

\bibitem[\protect\citeauthoryear{{Br{\" u}ggen}}{{Br{\"
  u}ggen}}{2003}]{brueggen03}
{Br{\" u}ggen} M.,  2003, ApJ, 593, 700

\bibitem[\protect\citeauthoryear{{Br{\" u}ggen} \& {Kaiser}}{{Br{\" u}ggen} \&
  {Kaiser}}{2002}]{nature}
{Br{\" u}ggen} M.,  {Kaiser} C.~R.,  2002, Nat., 418, 301

\bibitem[\protect\citeauthoryear{{Carilli} \& {Taylor}}{{Carilli} \&
  {Taylor}}{2002}]{carilli02}
{Carilli} C.~L.,  {Taylor} G.~B.,  2002, ARA\&A, 40, 319

\bibitem[\protect\citeauthoryear{{Cho}, {Lazarian}, {Honein}, {Knaepen},
  {Kassinos} \& {Moin}}{{Cho} et~al.}{2003}]{cho03}
{Cho} J.,  {Lazarian} A.,  {Honein} A.,  {Knaepen} B.,  {Kassinos} S.,
  {Moin} P.,  2003, ApJ, 589, L77

\bibitem[\protect\citeauthoryear{{Choudhuri}}{{Choudhuri}}{1998}]{plasmas}
{Choudhuri} A.,  1998, The Physics of Fluids and Plasmas, an introduction for
  astrophyisicists.
Cambridge University Press

\bibitem[\protect\citeauthoryear{{Churazov}, {Br{\" u}ggen}, {Kaiser}, {B{\"
  o}hringer} \& {Forman}}{{Churazov} et~al.}{2001}]{bub01}
{Churazov} E.,  {Br{\" u}ggen} M.,  {Kaiser} C.~R.,  {B{\" o}hringer} H.,
  {Forman} W.,  2001, ApJ, 554, 261

\bibitem[\protect\citeauthoryear{{Cowie} \& {McKee}}{{Cowie} \&
  {McKee}}{1977}]{cowie77}
{Cowie} L.~L.,  {McKee} C.~F.,  1977, ApJ, 211, 135

\bibitem[\protect\citeauthoryear{{Dalla Vecchia}, {Bower}, {Theuns}, {Balogh},
  {Mazzotta} \& {Frenk}}{{Dalla Vecchia} et~al.}{2004}]{vecchia04}
{Dalla Vecchia} C.,  {Bower} R.~G.,  {Theuns} T.,  {Balogh} M.~L.,  {Mazzotta}
  P.,    {Frenk} C.~S.,  2004, MNRAS, 355, 995

\bibitem[\protect\citeauthoryear{{Dolag}, {Jubelgas}, {Springel}, {Borgani} \&
  {Rasia}}{{Dolag} et~al.}{2004}]{dolag04}
{Dolag} K.,  {Jubelgas} M.,  {Springel} V.,  {Borgani} S.,    {Rasia} E.,
  2004, ApJ, 606, L97

\bibitem[\protect\citeauthoryear{{Edge}}{{Edge}}{2001}]{edge01}
{Edge} A.~C.,  2001, MNRAS, 328, 762

\bibitem[\protect\citeauthoryear{{Edge} \& {Stewart}}{{Edge} \&
  {Stewart}}{1991}]{edge91}
{Edge} A.~C.,  {Stewart} G.~C.,  1991, MNRAS, 252, 414

\bibitem[\protect\citeauthoryear{{Edge}, {Stewart} \& {Fabian}}{{Edge}
  et~al.}{1992}]{edgerb92}
{Edge} A.~C.,  {Stewart} G.~C.,    {Fabian} A.~C.,  1992, MNRAS, 258, 177

\bibitem[\protect\citeauthoryear{{Ettori} \& {Fabian}}{{Ettori} \&
  {Fabian}}{2000}]{ettori00}
{Ettori} S.,  {Fabian} A.~C.,  2000, MNRAS, 317, L57

\bibitem[\protect\citeauthoryear{{Evrard}}{{Evrard}}{1990}]{evrard90}
{Evrard} A.~E.,  1990, ApJ, 363, 349

\bibitem[\protect\citeauthoryear{{Fabian}}{{Fabian}}{1994}]{fab94}
{Fabian} A.~C.,  1994, ARA\&A, 32, 277

\bibitem[\protect\citeauthoryear{{Fabian}, {Allen}, {Crawford}, {Johnstone},
  {Morris}, {Sanders} \& {Schmidt}}{{Fabian} et~al.}{2002}]{missing02}
{Fabian} A.~C.,  {Allen} S.~W.,  {Crawford} C.~S.,  {Johnstone} R.~M.,
  {Morris} R.~G.,  {Sanders} J.~S.,    {Schmidt} R.~W.,  2002, MNRAS, 332, L50

\bibitem[\protect\citeauthoryear{{Fryxell}, {Olson}, {Ricker}, {Timmes},
  {Zingale}, {Lamb}, {MacNeice}, {Rosner}, {Truran} \& {Tufo}}{{Fryxell}
  et~al.}{2000}]{fryxell}
{Fryxell} B.,  {Olson} K.,  {Ricker} P.,  {Timmes} F.~X.,  {Zingale} M.,
  {Lamb} D.~Q.,  {MacNeice} P.,  {Rosner} R.,  {Truran} J.~W.,    {Tufo} H.,
  2000, ApJ Supp., 131, 273

\bibitem[\protect\citeauthoryear{{Ghizzardi}, {Molendi}, {Pizzolato} \& {De
  Grandi}}{{Ghizzardi} et~al.}{2004}]{ghizzardi04}
{Ghizzardi} S.,  {Molendi} S.,  {Pizzolato} F.,    {De Grandi} S.,  2004, ApJ,
  609, 638

\bibitem[\protect\citeauthoryear{{Kaiser}, {Pavlovski}, {Pope} \&
  {Fangohr}}{{Kaiser} et~al.}{2005}]{instab05}
{Kaiser} C.~R.,  {Pavlovski} G.,  {Pope} E.~C.~D.,    {Fangohr} H.,  2005,
  ArXiv Astrophysics e-prints

\bibitem[\protect\citeauthoryear{{Loeb}}{{Loeb}}{2002}]{loeb02}
{Loeb} A.,  2002, New Astronomy, 7, 279

\bibitem[\protect\citeauthoryear{{Malyshkin} \& {Kulsrud}}{{Malyshkin} \&
  {Kulsrud}}{2001}]{malyshkin01}
{Malyshkin} L.,  {Kulsrud} R.,  2001, ApJ, 549, 402

\bibitem[\protect\citeauthoryear{{Markevitch}, {Mazzotta}, {Vikhlinin},
  {Burke}, {Butt}, {David}, {Donnelly}, {Forman}, {Harris}, {Kim}, {Virani} \&
  {Vrtilek}}{{Markevitch} et~al.}{2003}]{markevitch03}
{Markevitch} M.,  {Mazzotta} P.,  {Vikhlinin} A.,  {Burke} D.,  {Butt} Y.,
  {David} L.,  {Donnelly} H.,  {Forman} W.~R.,  {Harris} D.,  {Kim} D.-W.,
  {Virani} S.,    {Vrtilek} J.,  2003, ApJ, 586, L19

\bibitem[\protect\citeauthoryear{{Narayan} \& {Medvedev}}{{Narayan} \&
  {Medvedev}}{2001}]{narmed01}
{Narayan} R.,  {Medvedev} M.~V.,  2001, ApJ, 562, L129

\bibitem[\protect\citeauthoryear{{Navarro}, {Frenk} \& {White}}{{Navarro}
  et~al.}{1995}]{nfw95}
{Navarro} J.~F.,  {Frenk} C.~S.,    {White} S.~D.~M.,  1995, MNRAS, 275, 56

\bibitem[\protect\citeauthoryear{{Navarro}, {Frenk} \& {White}}{{Navarro}
  et~al.}{1997}]{nfw97}
{Navarro} J.~F.,  {Frenk} C.~S.,    {White} S.~D.~M.,  1997, ApJ, 490, 493

\bibitem[\protect\citeauthoryear{{Nipoti} \& {Binney}}{{Nipoti} \&
  {Binney}}{2004}]{NB04}
{Nipoti} C.,  {Binney} J.,  2004, MNRAS, 349, 1509

\bibitem[\protect\citeauthoryear{{Ruszkowski} \& {Begelman}}{{Ruszkowski} \&
  {Begelman}}{2002}]{ruszbegel02}
{Ruszkowski} M.,  {Begelman} M.~C.,  2002, ApJ, 581, 223

\bibitem[\protect\citeauthoryear{{Rybicki} \& {Lightman}}{{Rybicki} \&
  {Lightman}}{1979}]{rybicki}
{Rybicki} G.~B.,  {Lightman} A.~P.,  1979, Radiative Processes in Astrophysics.
Wiley-Interscience

\bibitem[\protect\citeauthoryear{{Sarazin}}{{Sarazin}}{1986}]{sarazin}
{Sarazin} C.~L.,  1986, X-ray Emission from Clusters of Galaxies.
Reviews of Modern Physics

\bibitem[\protect\citeauthoryear{{Spitzer}}{{Spitzer}}{1962}]{spitzer}
{Spitzer} L.,  1962, Physics of Fully Ionized Gases.
Wiley-Interscience, New York

\bibitem[\protect\citeauthoryear{{Sutherland} \& {Dopita}}{{Sutherland} \&
  {Dopita}}{1993}]{sutherland}
{Sutherland} R.,  {Dopita} M.,  1993, ApJ Supp., 88, 253

\bibitem[\protect\citeauthoryear{{Tabor} \& {Binney}}{{Tabor} \&
  {Binney}}{1993}]{tabor93}
{Tabor} G.,  {Binney} J.,  1993, MNRAS, 263, 323

\bibitem[\protect\citeauthoryear{{Tribble}}{{Tribble}}{1989}]{tribble89}
{Tribble} P.~C.,  1989, MNRAS, 238, 1247

\bibitem[\protect\citeauthoryear{{Vikhlinin}, {Markevitch}, {Forman} \&
  {Jones}}{{Vikhlinin} et~al.}{2001}]{Vik2001}
{Vikhlinin} A.,  {Markevitch} M.,  {Forman} W.,    {Jones} C.,  2001, ApJ, 555,
  L87

\bibitem[\protect\citeauthoryear{{Voigt} \& {Fabian}}{{Voigt} \&
  {Fabian}}{2004}]{voigt04}
{Voigt} L.~M.,  {Fabian} A.~C.,  2004, MNRAS, 347, 1130

\bibitem[\protect\citeauthoryear{{Voigt}, {Schmidt}, {Fabian}, {Allen} \&
  {Johnstone}}{{Voigt} et~al.}{2002}]{voigt02}
{Voigt} L.~M.,  {Schmidt} R.~W.,  {Fabian} A.~C.,  {Allen} S.~W.,
  {Johnstone} R.~M.,  2002, MNRAS, 335, L7

\bibitem[\protect\citeauthoryear{{Woodward} \& {Colella}}{{Woodward} \&
  {Colella}}{1984}]{woodward}
{Woodward} P.,  {Colella} P.,  1984, JCP, 54, 174

\bibitem[\protect\citeauthoryear{{Zakamska} \& {Narayan}}{{Zakamska} \&
  {Narayan}}{2003}]{zakamska03}
{Zakamska} N.~L.,  {Narayan} R.,  2003, ApJ, 582, 162

\end{thebibliography}

\appendix

\section{Numerical Resolution Convergence} \label{app:numres}
The numerical convergence of the simulations was tested by running
simulation 0.01$ \kappa $ with varied spatial resolution. We compare
the evolution of temperature and density profiles for otherwise
identical simulations in which the minimum refinement, maximum
refinement and region of increased initial resolution in the central
region are all increased by one level of refinement. For the size of
the computational box we employ for these simulations, a refinement
level of 9 corresponds to a resolution of 0.31 kpc; a refinement level
of 10 corresponds to 0.15 kpc. The initial enhanced resolution of the
central region (refinement of 6) corresponds to 2.53 kpc in the
central $\sim $16 kpc. We also investigate the effect of increasing
this resolution by one level of refinement to a resolution of 1.26
kpc. The resolution in the outer regions of the cluster is determined
by the minimum refinement level. This is initially set to 3
(corresponding to roughly 20 kpc), but is also tested at level 4
(roughly 10 kpc).  The refinement parameters corresponding to each
simulation are summarised in table A1.

\begin{table*}
\centering
\begin{minipage}{110mm}
\caption{Summary of the refinement parameters of the numerical
resolution tests.}
\begin{tabular}{ccccc} \hline
 Test  & Central Refinement & Maximum refinement & Minimum refinement\\ 
\hline 
1 &  7 & 9 & 3\\ 
\hline
2 & 6 & 9 & 3\\
 \hline 
3 & 7 & 10 & 3\\ 
\hline 
4 & 8 & 9 & 3\\
 \hline
5 & 7 & 9 & 4\\
\hline
\end{tabular}
\end{minipage}
\end{table*}

\begin{figure*}
\centering \includegraphics[width=11cm]{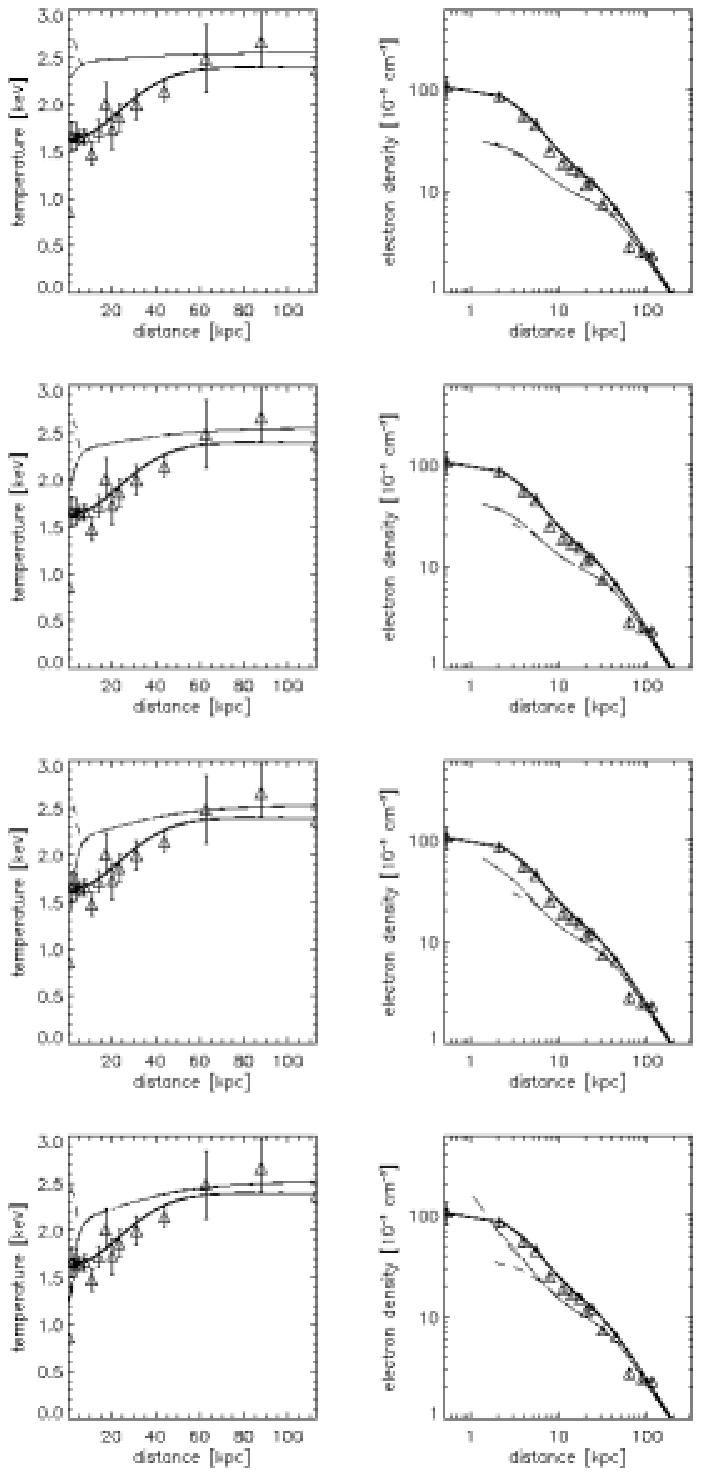}
\caption{\label{fig:restest}Comparison of simulation 0.01$ \kappa $
temperature and density profiles at each of the different numerical
resolutions at intervals of $3.16\times 10^{8}$ yrs starting at
$3.16\times 10^{8}$ yrs. The results for test 1 are shown by
dot dashed lines, test 2: dashed, test 3: triple dot dash, test 4:
long dash, test 5: dot}
\end{figure*}

From figure \ref{fig:restest} it is evident that only for a central
resolution of refinement level 6, or lower, do we observe
significantly different results to any other simulation. In this
particular case it appears that the central resolution is not
sufficient for the inner regions to be in hydrostatic equilibrium when
the simulation is initialised. This results in a slight collapse of
material near the cluster centre and therefore significant
compressional heating. For the other simulations the differences in
the temperature and density plots are minimal. The central effective
adiabatic index is also unchanged and in good agreement across the
range of different resolutions. This series of tests demonstrates that
for a central refinement of level 7, a maximum refinement of level 9
and a minimum refinement of level 3 the results are in close agreement
with those achieved with higher resolutions.  

\section{Hydrodynamic
Differences between 3-d and 1-d coordinate systems geometries}
\label{app:1d3d}

In this section we outline the differences between a purely
spherically symmetric geometry and a full 3-d geometry, for our
problem, with the aid of the hydrodynamic equations and trial
simulations of the Virgo cluster. For these purposes we have re-done
two of the simulations from the main paper with i) zero thermal
conductivity and viscosity and ii) full Spitzer thermal conductivity
and viscosity, in 1-d spherical coordinates, both at the same spatial
resolution as our 3-d runs and higher spatial resolutions. The minimum
level of refinement for the 1-d simulation performed at the same
spatial resolution as the 3-d runs was 7 corresponding to roughly 1.27
kpc. For the higher spatial resolution 1-d test the minimum refinement
was level 9, corresponding to a spatial scale of nearly 0.32 kpc. In
both cases the maximum level of the refinement was 12 corresponding to
approximately 0.04 kpc.

\begin{figure*}
\centering \includegraphics[width=14cm]{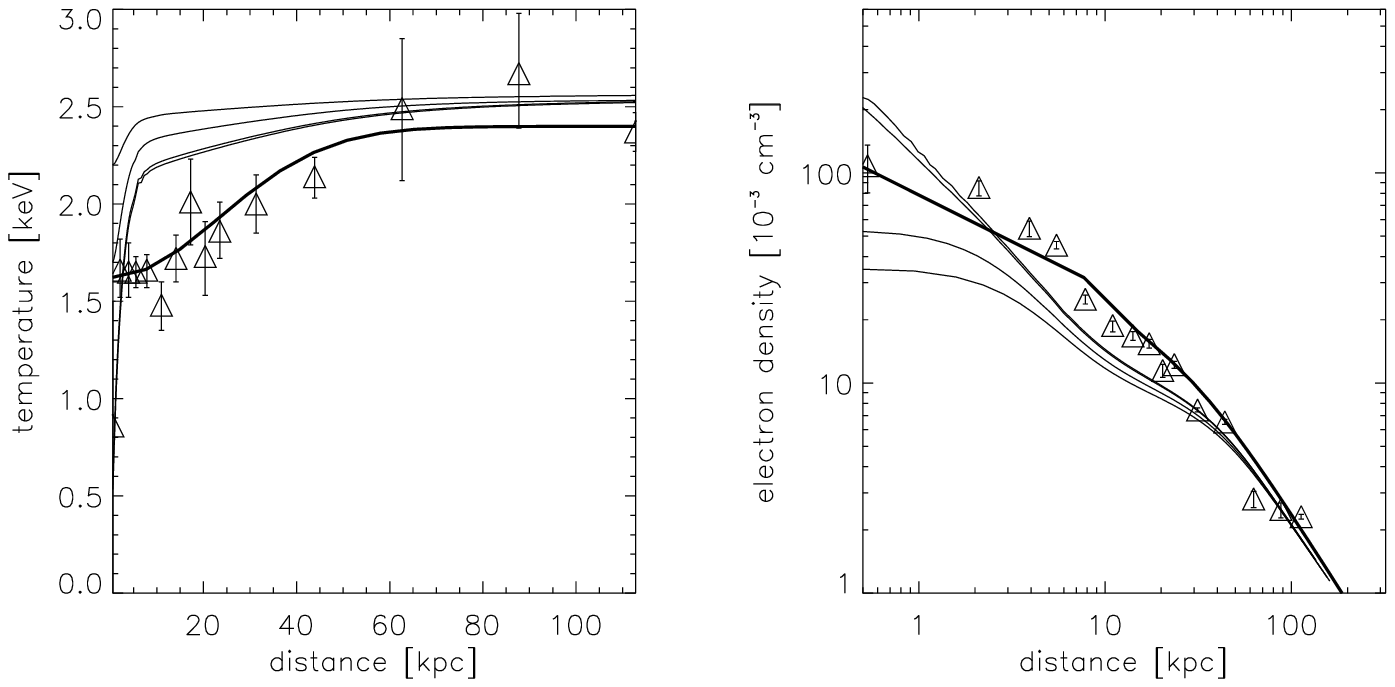}
\caption{Temperature and density profiles evolving with time for
simulation 0$ \kappa $ for a 1-d spherically symmetric geometry. The
profiles are plotted for the same times as the equivalent temperature
and density profiles in section \ref{res1}, with the same spatial
resolution, except for the final profile which is plotted at $1 \times
10^{9} $yrs.}
\label{fig:1dsph0thspitz}
\end{figure*}

\begin{figure*}
\centering \includegraphics[width=14cm]{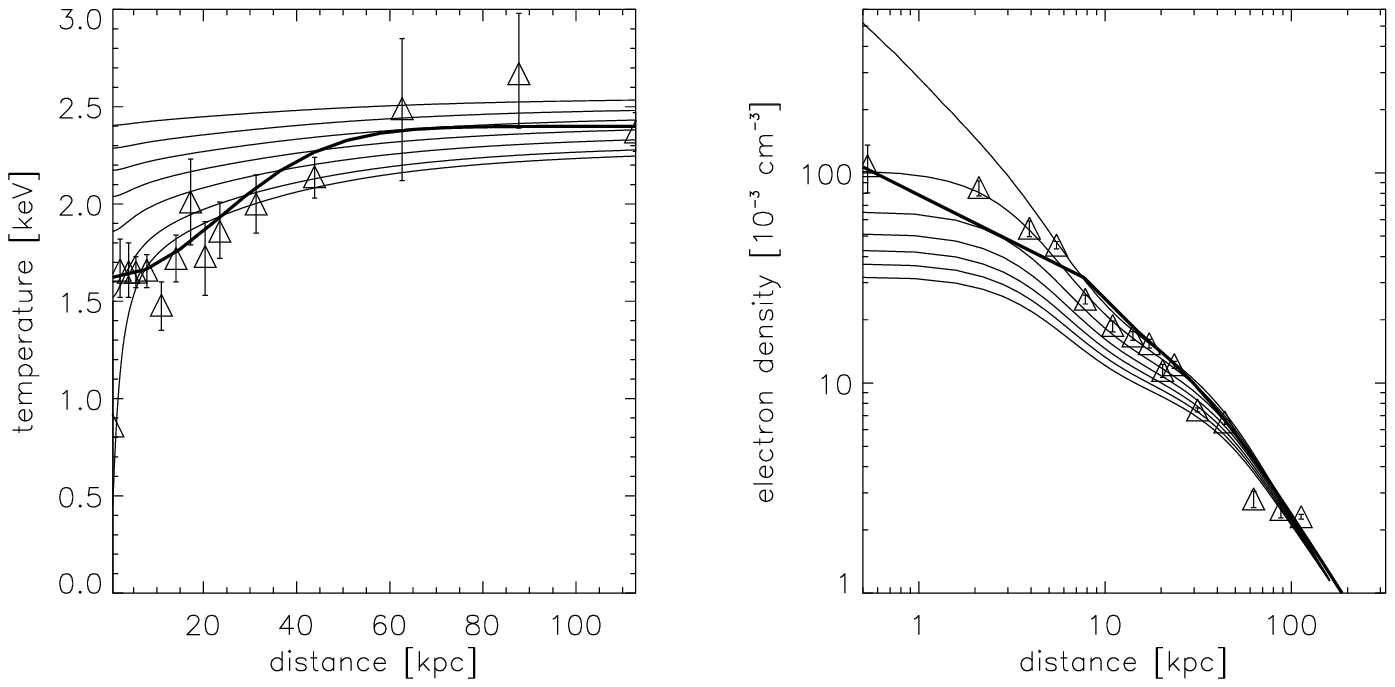}
\caption{Temperature and density profiles evolving with time for
simulation 1$ \kappa $ for a 1-d spherically symmetric geometry. The
profiles are plotted for the same times as the equivalent temperature
and density profiles in section \ref{res1}, with the same spatial
resolution, except for the final profile which is plotted at $4.2
\times 10^{9} $yrs.}
\label{fig:1dsph1thspitz}
\end{figure*}

The results of these 1-d spherically symmetric simulations explicitly
show that a cooling catastrophe always occurs in 1-d as well as
3-d. Furthermore, the cooling catastrophe takes less time to evolve in
the 1-d spherically symmetric case than the 3-d case and increasing
spatial resolution has almost no effect. For the simulation with zero
thermal conductivity the cooling catastrophe occurs after
approximately $1 \times 10^{9} $yrs-roughly $3 \times 10^{8}$ yrs
before the occurance in 3-d. For the simulation with full Spitzer
thermal conductivity the cooling catastrophe occurs after roughly $4.2
\times 10^{9} $yrs which is approximately $0.5 \times 10^{9} $yrs less
than in the 3-d case (see figures \ref{fig:1dsph0thspitz} and
\ref{fig:1dsph1thspitz}).

\begin{figure*}
\centering \includegraphics[width=14cm]{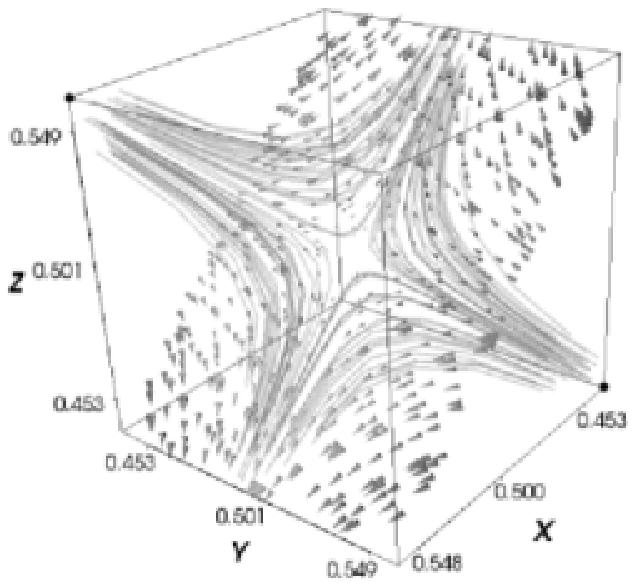}
\caption{Zoom on the centre of the simulation box. Velocity cut plane
(velocity vectors shown as cones) highlights the total velocity
vectors, demonstrating the essential non-radial features of the flow.
We have included two sets of particle tracers (grey flux tubes)
originating from the opposite corners (first set of 30 particle
tracers has centre at (0.450,0.550,0.450) and the second set at
(0.550,0.450,0.550)). The coordinates are shown in fraction of total
box width ($1. \times 10^{24}$ cm).}
\label{fig:3dflow}
\end{figure*}

Figure \ref{fig:3dflow} gives an indication of the non-radial flow in
the central regions of our simulated clusters, in the absence of
thermal conduction and viscosity. If the differences for simulations
using different geometries are real physical effects, rather than
numerical errors, then we should also expect to see differences in the
equations of hydrodynamics for the 1-d and 3-d cases. Since viscosity
enters into the momentum equation it is an appropriate choice to
investigate any differences which may arise due to dimensional
effects.

The equation for the viscous force per unit volume is given by
\citep[e.g.][]{sarazin}

\begin{equation}\label{eq:Fvis}
\bm {F}_{\rm vis} = \eta \bigg(\nabla^{2}\bm{v} + \frac{1}{3}\bm
{\nabla \nabla} . \bm{v} \bigg),
\end{equation}
where $\eta$ is the bulk viscosity and $v$ is fluid velocity.

Using the standard vector identity we can substitute for the
divergence term
\begin{equation}\label{eq:id}
\nabla^{2} \bm{v} = \bm{\nabla} (\bm{\nabla} . \bm{v}) - \bm{\nabla}
\times (\bm{\nabla} \times \bm{v} )
\end{equation}
so that equation (\ref{eq:Fvis}) becomes
\begin{equation}\label{eq:Fvis2}
\bm{F}_{\rm vis} = \eta \bigg(\frac{4}{3}\nabla^{2}\bm{v} +
\frac{1}{3}\bm{\nabla} \times (\bm{\nabla} \times \bm{v} )\bigg).
\end{equation}

Even without the evidence from simulations, from equation
(\ref{eq:Fvis2}) alone it is evident that viscous processes are
different in 3-d and 1-d cases due to shear effects. In 3-dimensions
the term involving the vector products constitutes a mechanism for the
dissipation of momentum, whereas in 1-d all vector products are zero,
by definition, and so this mechanism is suppressed.

It seems that equation (\ref{eq:Fvis2}) is able to account for some of
the differences between the 1-d and 3-d simulations in which full
Spitzer thermal conductivity and viscosity were present. However,
there was also a significant difference in the time at which the
cooling catastrophe occured in the simulations in which diffusion
processes were not present. Studying the complete momentum equation
allows us to investigate these differences.

\begin{equation}\label{eq:momentum}
\frac{\partial \bm{v}}{\partial t} + \big(\bm{v}. \bm{\nabla} \big)
\bm{v} = \bm{F} - \frac{1}{\rho} \bm{\nabla} P +\nu
\bigg(\frac{4}{3}\nabla^{2}\bm{v} + \frac{1}{3}\bm{\nabla} \times
(\bm{\nabla} \times \bm{v} )\bigg),
\end{equation}
where $F$ is the force per unit mass, $P$ is the fluid pressure and
$\nu$ is the kinematic viscosity.

Using a second standard vector calculus identity for the second term
on the left hand side of equation (\ref{eq:momentum}) we find
\begin{equation}\label{eq:id2}
\big(\bm{v}. \bm{\nabla} \big) \bm{v} = \frac{1}{2}\bm{\nabla}
(\bm{v}.\bm{v}) - \bm{v} \times (\bm{\nabla} \times \bm{v} ),
\end{equation}
and substituting into equation (\ref{eq:momentum}), the total momentum
equation is
\begin{equation}\label{eq:momentum2}
\frac{\partial \bm{v}}{\partial t} + \frac{1}{2}\nabla (\bm{v}.\bm{v})
- \bm{v} \times (\bm{\nabla} \times \bm{v} ) = \bm{F} - \frac{1}{\rho}
\bm{\nabla} P +\nu \bigg(\frac{4}{3}\nabla^{2}\bm{v} +
\frac{1}{3}\bm{\nabla} \times (\bm{\nabla} \times \bm{v} )\bigg).
\end{equation}

Since vector products are zero in 1-dimension, equation
(\ref{eq:momentum2}) reduces to
\begin{equation}\label{eq:momentum1d}
\frac{\partial v}{\partial t} + \frac{1}{2}\frac{\partial
(v^{2})}{\partial r} = F - \frac{1}{\rho} \frac{\partial P}{\partial
r} +\nu \frac{4}{3}\frac{1}{r^{2}}\frac{\partial}{\partial r}
\bigg(r^{2} \frac{\partial v}{\partial r}\bigg),
\end{equation}
in 1-d spherically symmetric coordinates.

From equation (\ref{eq:momentum1d}) it is clear that there is an
additional dissipation term in the 3-d case which is not present in
1-d, even in the absence of viscosity. Thus, in agreement with our
simulations, it is reasonable to expect the behaviour of a fluid to be
different in 1-d compared to 3-d, even in the absence of viscosity.

We note that the overall properties of the simulations, e.g. the
temperature and density profiles, remain roughly the same in 1-d
spherically symmetric and 3-d coordinates. Therefore, for simulations
such as these it seems possible to gain a reasonable estimate of the
main properties and timescales for the given circumstances at a lower
computational expense than for 3-d simulations. However, in order to
gain a more complete understanding of the physical processes at work
it is necessary to perform even these basic simulations in 3-d due to
the non-spherically symmetric process which may occur in both viscous
and inviscid fluid dynamics.

\label{lastpage} 
\end{document}